\documentclass[aps,prb,reprint,superscriptaddress,superscriptreference]{revtex4-1}

\usepackage{amsmath,dsfont,bm,amssymb,scalerel}
\usepackage{amsmath,bm,amssymb}
\usepackage{graphicx}
\usepackage{color}
\usepackage{hyperref}
\usepackage{enumitem}
\usepackage{algorithm,algpseudocode}
\usepackage{multirow}
\usepackage{colortbl,booktabs}
\makeatletter
\newcommand{\algmargin}{\the\ALG@thistlm}
\makeatother
\algnewcommand{\parState}[1]{\State%
  \parbox[t]{\dimexpr0.9\linewidth-\algmargin}{\strut #1\strut}
}
\algnewcommand{\Inputs}[1]{%
  \State \textbf{Inputs:}
  \Statex \hspace*{\algorithmicindent}\parbox[t]{.8\linewidth}{\raggedright #1}
}
\algnewcommand{\Initialize}[1]{%
  \State \textbf{Initialize:}
  \Statex \hspace*{\algorithmicindent}\parbox[t]{.8\linewidth}{\raggedright #1}
}

\AtBeginDocument{\nocite{achemso-control}}
\newcommand*{\citen}{}
\DeclareRobustCommand*{\citen}[1]{%
	\begingroup
	\romannumeral-`\x 
	\setcitestyle{numbers}%
	\cite{#1}%
	\endgroup
}

\newcommand\equalhat{%
	\let\savearraystretch\arraystretch
	\renewcommand\arraystretch{0.3}
	\begin{array}{c}
		\stretchto{
			\scalerel*[\widthof{=}]{\wedge}
			{\rule{1ex}{3ex}}%
		}{0.5ex}\\ 
		=%
	\end{array}
	\let\arraystretch\savearraystretch
}
\newcommand{\myrowcolour}{\rowcolor[gray]{0.925}}

\newcommand{\gj}[1]{\textcolor{black}{#1}}
\newcommand{\fs}[1]{\textcolor{black}{#1}}

\newcommand{\ud}{\text{d}}
\newcommand{\kB}{k_\text{B}}
\newcommand{\Dt}[1][]{\Delta t_\text{#1}}
\newcommand{\rcut}[1][]{r_\text{c#1}}

\newcommand{\RR}{\bm{R}}
\newcommand{\VV}{\bm{V}}
\newcommand{\FF}{\bm{F}}
\newcommand{\FFC}[1][]{\bm{F}_{#1}^\text{C}}
\newcommand{\FFD}[1][]{{\bm{F}_{#1}^\text{D}}'}
\newcommand{\FFR}[1][]{\bm{F}_{#1}^\text{R}}
\newcommand{\KK}{\bm{K}}
\newcommand{\KKself}[1][]{\bm{K}_{#1}^{\text{self}}}
\newcommand{\KKs}[1][]{\bm{K}_{#1}^{\text{s}}}
\newcommand{\DKKs}[1][]{\Delta \bm{K}_{#1}^{\text{s}}}
\newcommand{\KKpair}[1][]{\bm{K}_{#1}^{\text{p}}}
\newcommand{\Ks}[1][]{{K}_{#1}^{\text{s}}}
\newcommand{\DKs}[1][]{\Delta {K}_{#1}^{\text{s}}}
\newcommand{\Kpair}[1][]{{K}_{#1}^{\text{p}}}
\newcommand{\mmax}{m_\text{max}}
\newcommand{\bbb}{\bm{b}}
\newcommand{\aaa}{\bm{a}}
\newcommand{\Full}[1]{\mathcal{#1}}
\newcommand{\Fourier}[1]{\hat{\mathcal{#1}}_{\omega}}
\newcommand{\Ca}{C^\text{a}}
\newcommand{\Cc}{C^\text{c}}

\newcommand{\DPC}[1][]{\Delta \bm{P}_{#1}^C}
\newcommand{\DPD}[1][]{\Delta {\bm{P}_{#1}^D}'}
\newcommand{\DPDpp}[1][]{\Delta {\bm{P}_{#1}^D}''}
\newcommand{\DPR}[1][]{\Delta \bm{P}_{#1}^R}

\begin{document}

\title{Generalized Langevin dynamics: Construction and numerical integration 
of non-Markovian particle-based models}

\author{Gerhard Jung}
\email{jungge@uni-mainz.de}
\affiliation{Institut f\"ur Physik, Johannes Gutenberg-Universit\"at Mainz, 
Staudingerweg 9, 55128 Mainz, Germany}
\affiliation{Graduate School of Excellence Materials Science in Mainz, Staudingerweg 9, 55128 Mainz, Germany}
\author{Martin Hanke}
\email{hanke@mathematik.uni-mainz.de}
\affiliation{Institut f\"ur Mathematik, Johannes Gutenberg-Universit\"at Mainz, 
	Staudingerweg 9, 55128 Mainz, Germany}
\author{Friederike Schmid}
\email{friederike.schmid@uni-mainz.de}
\affiliation{Institut f\"ur Physik, Johannes Gutenberg-Universit\"at Mainz, 
Staudingerweg 9, 55128 Mainz, Germany}

\begin{abstract}

We propose a \gj{Generalized Langevin dynamics} (GLD) technique to construct
non-Markovian particle-based coarse-grained models from fine-grained reference
simulations and to efficiently integrate them. The proposed \gj{GLD} model has the
form of a discretized generalized Langevin equation with distance-dependent two-particle
contributions to the self- and pair-memory kernels. The memory kernels are
iteratively reconstructed from the dynamical correlation functions of an
underlying fine-grained system. We develop a simulation algorithm for this
class of non-Markovian models that scales linearly with the number of
coarse-grained particles. Our \gj{GLD} method is suitable for coarse-grained studies
of systems with incomplete time scale separation, as is
often encountered, e.g., in soft matter systems.


We apply the method to a suspension of nanocolloids with frequency-dependent
hydrodynamic interactions. We show that the results from \gj{GLD} simulations
perfectly reproduce the dynamics of the underlying fine-grained system. The
effective speedup of these simulations amounts to a factor of about $ 10^4 $.
Additionally, the transferability of the coarse-grained model with respect to
changes of the nanocolloid density is investigated. {The results indicate
that the model is transferable to systems with {nanocolloid densities} that
{differ by up to one order of magnitude} from the density of the reference
system. }

\end{abstract}

\maketitle


\section{Introduction}

Many materials, in particular soft matter, exhibit dynamics on multiple length
and time scales. Examples are protein folding \cite{FrembgenKesner2009},
polymer aggregation \cite{Tomilov2013} or colloidal crystallization
\cite{Palberg1999,Vermant2005}. Since these problems are difficult to access
with standard atomistic simulations, one typically considers coarse-grained
(CG) models with fewer degrees of freedom. A vast variety of different
coarse-graining techniques have been suggested that reduce the dimensionality
of a system \fs{\cite{Ermak1975, VanGunsteren1982, Lyubartsev1995, Reith2003,
Izvekov2004, Shell2008,Hijon2010}}.  Most of these methods either do not target
the dynamical properties of the fine-grained system at all, or they assume
complete time scale separation: They postulate that the dynamics of the
relevant, coarse-grained particles is much slower than the relaxation time
of the irrelevant, neglected degrees of freedom, and thus take the CG dynamics
to be Markovian. In real systems, however, the relevant time scales often
overlap and the Markovian assumption is not justified. In such cases, a more
appropriate framework for the construction of dynamic coarse-grained models is
the (multidimensional) generalized Langevin equation (GLE) \cite{Kinjo2007}
\begin{equation}
 M \dot{\VV}_i(t) 
 = \FFC[i](t) - \int_{0}^{t} \!\!\!\! \ud s \: \sum_j \KK_{ij}(t,s) \VV_j(s) 
    + \partial \FF_i(t),
\label{eq:GLE}
\end{equation}
where $\VV_i(t)$ denotes the velocity of the CG particle $i$, $M$ its mass,
$\FFC[i](t)$ the conservative forces, and $\partial \FF_i(t)$ the fluctuating
forces acting on it. The memory kernel tensor $\KK_{ij}(t,s)$ determines the
non-Markovian dissipative self- and pair-interactions of the CG particles. Note
that both $\FFC[i]$ and $\KK_{ij}$ are functions of the positions $\RR_j$ of
all particles $j$. The fluctuating force is related to the memory kernel
tensor via the fluctuation-dissipation theorem (FDT),
\begin{equation}
\left\langle \partial\FF_i(t) \partial\FF_j(t') \right\rangle 
  = \kB T \: \KK_{ij} (t,t'),
\label{eq:FDT}
\end{equation}
with the Boltzmann constant $ \kB $ and the thermodynamic temperature $ T $.
The GLE thus describes a system of particles in a canonical ensemble with
constant temperature $ T $.  The general form of the GLE results from the 
Mori-Zwanzig projection operator formalism, which  was introduced 
roughly 50 years ago to theoretically understand the process of 
systematic coarse-graining \cite{Zwanzig1961, Mori1965, Zwanzig2001}.
\fs{We note that  the average $\left\langle \cdot \right \rangle$ in
Eq.\ (\ref{eq:FDT}) is a conditional average for fixed CG configuration 
and thus depends on the position of {\em all} CG particles, just like the
memory kernel.}

In recent years, the GLE has become increasingly popular as a tool for
mesoscopic modeling \fs{\cite{Smith1990, Ceriotti2010, Cordoba2012,
Baczewski2013,Li2015,Davtyan2015, Li2017, Jung2017, Meyer2017}}. However, most
studies so far were restricted to systems {with} one or two particles. One
problem with GLE-modeling of many-particle systems is the high dimension of the
memory tensor $\KK_{ij}$, which complicates the generation of random forces
with correct correlations (Eq.\ (\ref{eq:FDT})). The problem can be avoided if
one simply neglects cross-correlations in the friction kernel ($\KK_{ij} =
\delta_{ij} \KK_{ii}$). An alternative Ansatz was recently proposed by Li
\emph{et al.} \cite{Li2015, Li2017}. These authors generalized the standard
dissipative particle dynamics (DPD) equations of motion to a non-Markovian DPD
(NM-DPD) method with frequency-dependent pair friction terms.  In NM-DPD, the
dissipative friction term in Eq.\ (\ref{eq:GLE}) is thus taken to have the form 
\begin{eqnarray}
\label{eq:NM-DPD}
\lefteqn{
\int_{0}^{t} \ud s \: \sum_{j} \KK_{ij}^{\text{NM-DPD}}{(t,s)} \VV_j(s)} 
\qquad \qquad
\\ \nonumber &&
= \int_{0}^{t} \ud s \: \sum_{j \neq i} \KK_{ij}(t-s) \: (\VV_j(s)-\VV_i(s)).
\end{eqnarray}
As in standard DPD, the correlated random forces $\partial \FF_i$ can then be
written as sums of uncorrelated random pair forces $\partial \FF_{ij} = -
\partial \FF_{ji}$, which greatly simplifies the problem of random force
generation.  

{To the best of our knowledge}, the method of Li \emph{et al.} is the only
method published so far that enables non-Markovian modeling with dissipative
pair-interactions.  They applied their formalism to star-polymer melts and
reported promising results. However, the NM-DPD assumption (\ref{eq:NM-DPD})
implies \fs{an additive relation} $\KK_{ii}^{\text{NM-DPD}} = - \sum_{j \neq i}
\KK_{ij}$ in Eq.\ (\ref{eq:GLE}), which is often not correct. Moreover,
NM-DPD models are Galilean invariant by construction, which is clearly
inappropriate if the CG model describes the motion of CG particles in a
background medium -- as is the case in implicit solvent models. 

In the present paper we propose a more general approach, the ''generalized
Langevin dynamics'' (GLD) method. Like the above-mentioned models, it is also
derived from the GLE, but it relies on weaker assumptions (the latter can in
fact be considered as special cases of the \gj{GLD} model). The method can be seen
as a generalization of Brownian dynamics (BD) with frequency-dependent friction
tensors \cite{Ermak1975, VanGunsteren1982}. It can therefore be applied to
polymers as well as colloidal systems with incomplete separation of time
scales. 

In a recent paper, we have investigated the dissipative pair-interactions in a
system of two isolated nanocolloids by theory and molecular dynamics
simulations \cite{Jung2017a}. Here, we consider suspensions of many
nanocolloids.  We construct a coarse-grained \gj{GLD} model from simulation data for
a fine-grained explicit solvent model at a reference nanocolloid density
$\rho_0$. Then we perform \gj{GLD} simulations at a set of densities $\rho$. This
allows us to validate the method (by comparing the results of the GLD
simulations at $\rho=\rho_0$ with those from the fine-grained simulations)
and to investigate other properties such as the transferability and
computational efficiency (benchmarks) of the \gj{GLD} method.

Our paper is organized as follows: In Sec.~\ref{sec:num_integration} we
introduce the \gj{Generalized Langevin dynamics} method. We derive the discretized
equations of motion and present a method how to efficiently determine the time-
and cross-correlated fluctuating forces. In Sec.~\ref{sec:reconstruction} we
then generalize the iterative memory reconstruction (IMRV) method published in
earlier work \cite{Jung2017} and demonstrate how it can be used to
reconstruct memory kernels from atomistic simulations. Our results are
presented and discussed in Sec.~\ref{sec:results}.  We summarize and conclude
in Sec.~\ref{sec:conclusion}.


\section{Generalized Langevin dynamics}
\label{sec:num_integration}

\subsection{Basic Model}
\label{ssec:GLD_basic}

In the following, we consider a system of $ N $ coarse-grained identical
particles with positions $ \bm{R}_i(t) $ and velocities $ \bm{V}_i(t) $ in
three dimensions. Our only approximation is to neglect dissipative
many-body interactions. The memory kernel in the generalized Langevin equation
for particle $ i $, Eq.\  (\ref{eq:GLE}), can then be written as
\begin{equation}
\KK_{ij}(t,s) = \left\{ \begin{array}{lcr}
  \KKself {[\{\RR_{ik}(t)\},t-s]} 
   &:& i = j \\
\KKpair {[\RR_{ij}(t),t-s]} &:& i \neq j
\end{array} \right. 
\label{eq:GLD_basic}
\end{equation}
with
\begin{equation}
  \KKself {[\{\RR_{ik}(t)\},\tau]}  =
  \KKs(\tau) + \sum_{k \neq i} \DKKs {[\RR_{ik}(t),\tau ]}
\label{eq:GLD_basic2}
\end{equation}
%
and $\RR_{ij}(t) = \RR_i(t) - \RR_j(t)$. Here we have expressed the memory
kernel tensor in terms of $3 \times 3$ dimensional ''pair-memory'' and ''self-memory'' kernels ($\KKpair$ and $\KKself$, respectively), and we have taken
into account the possibility that nearby particles may affect
the self-memory via the sum over $\DKKs$.  If the pair interactions are strong,
it is important to include the latter contribution to the self-interactions,
as has been shown in Ref.~\citen{Jung2017a} and will also be apparent in
Sec.~\ref{sec:reconstruction}. 

We expect that in many applications, the two-particle contributions to the
memory kernels, $\KKpair$ and $\DKKs$, can be separated into contributions
parallel and orthogonal to the line connecting the centers of the particles. If
this is the case, the two-particle contributions to the memory kernels can be
decomposed according to
\begin{eqnarray}
\KKpair {[\RR_{ij}(t),t']} & \approx &
  \Kpair[\parallel][R_{ij}(t),t'] \, \bm{{e}}_{ij}\bm{{e}}^T_{ij} 
\\ \nonumber
 &&+ \Kpair[\perp]{[R_{ij}(t),t']}(1-\bm{{e}}_{ij}\bm{{e}}^T_{ij} ),
\\ 
\DKKs{[\RR_{ij}(t),t']}  &\approx &
\DKs[\parallel][R_{ij}(t),t'] \, \bm{{e}}_{ij}\bm{{e}}^T_{ij} 
\\ \nonumber
 &&+ \DKs[\perp][R_{ij}(t),t'](1-\bm{{e}}_{ij}\bm{{e}}^T_{ij} ).
\end{eqnarray}
where $ R_{ij}(t) = |\RR_{ij}(t)| $ and $ \bm{{e}}_{ij} =
\RR_{ij}(t)/R_{ij}(t)$. In our numerical studies (Secs.\
\ref{sec:reconstruction} and \ref{sec:results}), we will only consider the
parallel components of $\KKpair$ and $\DKKs$ for
simplicity.  This approximation is justified by the theoretical and simulation
results in Ref.~\citen{Jung2017a}. Additionally, we will assume that the
single-particle memory kernel is isotropic, \fs{$\KKs(\tau)= \Ks(\tau)
\bm{1}$}, as it should be, given the symmetry of our system.

\subsection{Discretized Model Equations}
\label{ssec:GLD_discrete}

Having defined the basic model, Eq.~(\ref{eq:GLE}) with
{(\ref{eq:GLD_basic}) and (\ref{eq:GLD_basic2})}, our next task is to construct a
numerical integrator. To this end, we first introduce discretized memory 
kernels $\KK_m$, defined as
\begin{equation}
 \KK(t-s) = \sum_{m=0}^{\mmax-1} \KK_m \delta(t-m\Dt)
\label{eq:k-discrete}
\end{equation}
with the time step $ \Dt $. The cutoff $\mmax$ determines the longest time
scale $ \tau_\text{mem} = \mmax \Dt $ on which memory effects are considered in
the model. Introducing such a cutoff is necessary for numerical reasons. It
must be optimized such that the \gj{GLD} model is still computationally efficient
while capturing the relevant memory effects in the underlying microscopic
model. In some cases, the discretization and cutoff of the memory kernel can be
circumvented by introducing auxiliary variable expansions \cite{Ceriotti2010,
Cordoba2012, Cordoba2012a, Baczewski2013, Li2017}. Here, the idea is to replace
the non-Markovian equations of motion by Markovian equations for a system with
virtual additional degrees of freedom. The additional, auxiliary variables then
introduce memory effects in the dynamics of the ''real'' particles, and they
can be constructed in a systematic manner by fitting the target memory kernel
to a sum of (complex) exponentials (see Appendix \ref{sec:aux_var}).  For the
systems considered in the present work, it was, however, not possible to
utilize this expansion -- mainly due to problems with the distance-dependent
two-particle contributions to the memory, $\DKKs$ and $\KKpair$. This is
discussed in more detail in Appendix \ref{sec:aux_var}.   

Inserting the discrete version of the memory kernels, Eq.\
(\ref{eq:k-discrete}), the \gj{GLD} equation (\ref{eq:GLE}) with
{(\ref{eq:GLD_basic}) and (\ref{eq:GLD_basic2})} take the form
\begin{eqnarray}
\label{eq:GLE_discretized} 
M \dot{\VV}_i(t)&=& \FFC[i][\{\RR_j(t)\}] 
\\ \nonumber
&-& \sum_{m=0}^{\mmax-1} \bigg\{ 
  \sum_{j\neq i}  \KKpair[m][\RR_{ij}(t)] \: \VV_j(t-m\Dt)  
\\ \nonumber
&& \quad + \:
  \KKself[m][\{\RR_{ik}(t)\}] \: \VV_i(t-m \Dt) 
  \bigg\} + \partial \FF_i(t)
\end{eqnarray}
with the fluctuation-dissipation relation 
\begin{eqnarray}
\label{eq:FDT_discretized}
\lefteqn{\langle \partial \FF_i(t) \partial \FF_j(t') \rangle 
= \kB T \!\!\sum_{m=0}^{\mmax-1} \!\! a_m \delta(t-t'-m \Dt)} 
\quad \quad
 \\ \nonumber 
&&
\times
\Big( \delta_{ij} \KKself[m][\{\RR_{ik}(t)\}]
+ (1-\delta_{ij}) \KKpair[m][\RR_{ij}(t)] \Big),
\end{eqnarray}
where $ \KKself[m][\{\RR_{ik}\}] 
= \KKs[m] + \sum_{k \neq i} \DKKs[m][\RR_{ik}(t)]$
and the prefactors $a_m$ are given by $a_0 = 2$ and $a_m = 1$ for $m \neq 0$.
The derivation of these equations follows closely that of Eqs.~(12)-(17) in
Ref.~\citen{Jung2017} and will not be repeated here.

Starting from Eqs.~(\ref{eq:GLE_discretized})-(\ref{eq:FDT_discretized}),
we can derive a numerical integrator for the GLE following a scheme proposed
by Gr\o{}nbech-Jensen and Farago \cite{Gronbech-Jensen2012}.
In earlier work \cite{Jung2017}, we have applied this scheme to construct
an integrator for the single-particle GLE; here, we extend that work to 
multiparticle \gj{GLD} with two-particle contributions to the memory kernel. 
The derivation of the algorithm is presented in Appendix 
\ref{sec:GLD_integration}. The final equations read
\begin{eqnarray}
\label{eq:GLD_integrator}
\RR_{i,n+1} &=& \RR_{i,n}+ \Dt \: \aaa \: \VV_{i,n}
\\ \nonumber &&
    + \frac{\Dt^2}{2M} \: \bbb\: (\FFC[i,n] + \FFD[i,n]+ \FFR[i,n]) \\
\VV_{i,n+1} &=& 
   \aaa \: \VV_{i,n} + \frac{\Dt}{2M} \: (\aaa \: \FFC[i,n] + \FFC[i,n+1])
\\ \nonumber &&
    + \frac{\Dt}{M} \: \bbb \: (\FFD[i,n]+ \FFR[i,n]) 
\end{eqnarray}
where we have introduced the shortcut notations
\mbox{$\RR_{i,n} = \RR_{i}(n\Dt)$}, \mbox{$\VV_{i,n} = \VV_{i}(n\Dt)$},
\mbox{$\FFC[i,n] = \FFC[i][\{\RR_{j,n}\}]$}, and defined \gj{the matrices}
\begin{equation}
\label{eq:aa-bb}
\bbb = \Big[\bm{1}+\frac{\Dt}{2 M}\KKs[0]\Big]^{-1}, 
\quad
\aaa =  \bbb \: \Big[\bm{1}-\frac{\Dt}{2 M}\KKs[0]\Big],
\end{equation}
as well as \gj{the dissipative force vector}
\begin{eqnarray}
\label{eq:dissipative_force2}
\FFD[i,n] &=& 
  - \sum_{j \neq i} \big( \KKpair[0][\RR_{ij,n}]\: \VV_{j,n}
    + \DKKs[0][\RR_{ij,n}] \: \VV_{i,n} \big)
\\ \nonumber &&
- \frac{1}{\Dt} \!\!\! \sum_{m=1}^{\mmax-1} \bigg\{ 
    \sum_{j\neq i}  \KKpair[m][\RR_{ij,n}] \, \Delta\RR_{j,n-m}
\\ \nonumber &&
\qquad \qquad + \: 
    \Big( \KKs[m] + \sum_{k \neq i} \DKKs[m][\RR_{ik,n}] \Big) \,
     \Delta\RR_{i,n-m}  
       \bigg\} , 
\end{eqnarray}
with $\Delta \RR_{k,n} = \RR_{k,n+1}-\RR_{k,n}$.
$\FFR[i,n]$ are vectors of correlated random numbers with zero
mean and correlations given by
\begin{eqnarray}
\label{eq:stochastic_force2}
\langle \FFR[i,n+m] \FFR[j,n] \rangle &=& \kB T \frac{a_m}{\Dt}
\KKpair[m][\RR_{ij,n+m}], \quad \mbox{for} \: i \neq j 
\\ \nonumber 
\langle \FFR[i,n+m] \FFR[i,n] \rangle &=& \kB T \frac{a_m}{\Dt}
\Big( \KKs[m] + \sum_{k \neq i}\DKKs[m][\RR_{ik,n+m}] \Big).
\end{eqnarray}


The auto- and cross-correlations of the $3N $ dimensional stochastic force
vector can be described by a $(3N \times 3N)$ dimensional
correlation matrix (see below in Sec.~\ref{ssec:GLD_RNG}).  In principle, one
should prove that this matrix is positive definite for all possible
particle configurations. This was done, e.g., when establishing the
Rotne-Prager tensor as a useful mobility tensor in BD simulations with
hydrodynamic interactions \cite{Rotne1969}.  In practice, however, providing
such general proofs for numerically reconstructed memory kernels is very
challenging, therefore, we have checked the positive definiteness on-the-fly in
all simulations.

\subsection{Constructing the Stochastic Forces}
\label{ssec:GLD_RNG}

The remaining challenge in the development of our \gj{GLD} simulation  method is to
construct an algorithm that efficiently generates stochastic forces $\FFR[i,n]$
with correlations as prescribed by the fluctuation-dissipation theorem, Eq.\
(\ref{eq:stochastic_force2}). To this end, we generalize a technique
proposed by Barrat \emph{et al.} \cite{Barrat2011}, which was also
applied in Refs.~\citen{Li2015,Jung2017}. 

We first rewrite Eq.~(\ref{eq:stochastic_force2}) as a matrix equation,
\begin{equation}
 \left\langle \Full{F}_{n+m} \Full{F}_n \right\rangle = \Full{K}_m,
 \label{eq:correlation_matrix}
\end{equation}
where the entries $F_{I{,m}}$ of the $3N$ dimensional vectors $\Full{F}_m$ are the
unknown components {of the} time- and cross-correlated stochastic force vectors
$\FFR[i,m]$, and the known entries $K_{IJ{,m}}$ of the $3N\times 3N$ dimensional
correlation matrices $\Full{K}_m$ are determined by the right hand side of Eq.\
(\ref{eq:stochastic_force2}). 

Generalizing  Ref.\ \citen{Jung2017} in a straightforward manner, we introduce
the real and symmetric matrices $ \Full{A}_{s} $ for 
\mbox{$ s = -\mmax +1,...,\mmax-1 $},
\begin{equation}
\Full{K}_m \equiv \sum_{s=-\mmax+1}^{\mmax-1} \Full{A}_{s} \Full{A}_{s+m},
\label{eq:noise_ansatz}
\end{equation}
where $ \Full{A}_{s+m} =\Full{A}_{s+m-2\mmax+1} $ if $ s+m \geq \mmax $. The
stochastic force vector $ \Full{F}_{n} $ can now be calculated by multiplying
$ \Full{A}_{s} $ with a sequence of uncorrelated Gaussian
distributed random vectors $ \Full{W}_{n} $,
\begin{equation}
\Full{F}_{n} = \sum_{s=-\mmax+1}^{\mmax-1} \Full{A}_{s} \Full{W}_{n+s}.
\label{eq:noise_production}
\end{equation}

The time-consuming task is thus the determination of the matrices $\Full{A}_{s}
$. This is done by exploiting the convolution theorem. We Fourier transform 
the quantities $\Full{X}=\Full{F,W,K,A}$ according to
\begin{equation}
\Fourier{X} = \sum_{m=-\mmax+1}^{\mmax-1} \Full{X}_m 
\exp\left (-{\rm i}\omega m \frac{2 \pi }{2\mmax-1} \right),
\label{eq:FT}
\end{equation}
(taking $\Full{K}_{-m} = \Full{K}_m$) and insert this expression into
Eqs.\ (\ref{eq:noise_ansatz}) and (\ref{eq:noise_production}), which gives
\begin{equation}
\Fourier{K} = \Fourier{A} \Fourier{A}
\quad \mbox{and} \quad
\Fourier{F} = \Fourier{A} \Fourier{W} .
\end{equation}

This final result shows that we do not need to determine the matrix square root
$\Fourier{A} = \sqrt{\Fourier{K}}$ explicitly, but only the product of the
square root with the random input vector. Calculating the full square root
matrix scales with the third power of $ N $, while the latter can be realized
with linear scaling using the Lanczos algorithm
\cite{Hochbruck1997,Higham2008,Aune2013}, if one assumes that the particle
interactions have a cutoff radius $ \rcut$.  The basic idea behind the Lanczos
algorithm is to calculate the projection of a given $ n\times n $ dimensional
input matrix and a $ n $ dimensional vector onto a \emph{Krylov subspace}. This
is usually {applied} as an iterative method to determine the eigenvalues of the
input matrix. The outcome of the projection is a reduced symmetric, tridiagonal
matrix. For such a matrix, simple and fast techniques to determine the square
root are available \cite{numrec2007}. Therefore, the Lanczos algorithm can also
be used to determine the product of the input vector with the square root of
the input matrix.  

The pseudo code that {implements} the above described method is displayed in
Algorithm~\ref{alg:coloured_noise}.  In the following, we make some important
comments on the algorithm:

\begin{itemize}

	\item The bottleneck of the method is the matrix-vector multiplication
(steps 5,7,10). The results of steps 5 and 10 are thus stored and reused in
step 7.

	\item In our implementation of the algorithm, the fastest operation to
conduct the matrix-vector multiplication (steps 5,7) \gj{is to iterate over all neighbors
in every iteration step (e.g. by using neighbor lists)}. Using sparse-matrix multiplication was found to
be very time consuming.

	\item The square root of the matrix $ \bm{H}^{k+1} $ in step 13 can
simply be determined by exploiting the fact that $ \bm{H}^{k+1} $ is a real,
symmetric and tridiagonal matrix (see Ref.~\citen{numrec2007}).

\end{itemize}

\begin{algorithm}[H]
\caption{Generating correlated random numbers $ F_{I,n} $ with
the distribution $ \langle F_{I,n+m} F_{J,n} \rangle = K_{IJ,m}$}
 \begin{algorithmic}[1]
  \Inputs{
   $ K_{IJ,m} \text{ for } m = 0,...,\mmax-1 
              \text{ with } K_{IJ,m} = K_{IJ,-m} $\\
   $ W_{I,n} $ with $ \langle W_{I,n+m} W_{J,n} \rangle 
                     = \delta_{m0} \delta_{IJ} $ 
             	\vspace*{0.1cm} }
		
  \Initialize{
   compute $\hat{K}_{IJ,\omega} = \sum_{m=-\mmax+1}^{\mmax-1}  K_{IJ,m} 
           \exp(-{\rm i} m \omega \frac{2 \pi}{2\mmax-1})$\\
   compute $\hat{W}_{I,\omega} = \sum_{m=-\mmax+1}^{\mmax-1}  
           W_{I,n+m} \exp(-{\rm i} m \omega \frac{2 \pi}{2\mmax-1})$\\
		\vspace*{0.1cm} }
		
  \For{$ \omega = 0\text{ to }\mmax-1 $}
  \State set $ v^0_I = 0 $, $ \beta^0 = 0 $, 
         $ v^1_I = \hat{W}_{I,\omega} / \lVert \bm{\hat{W}}_\omega \rVert ,
         k=1,\Delta = 1$
  \State compute $ \alpha^1 = v^1_I \hat{K}_{IJ,\omega} v^1_J $
  \While{$ \Delta > tol $}
  \State compute $ r_I^{k+1} = \hat{K}_{IJ,\omega} v_J^{k} 
                   - \alpha^k v_I^{k} - \beta^{k-1}v_I^{k-1}$
  \State set $ \beta^k = \lVert \bm{r}^{k+1} \rVert  $
  \State set $ v_I^{k+1} = r_I^{k+1}/\beta^k  $
  \State compute $\alpha^{k+1} = v^{k+1}_I \hat{K}_{IJ,\omega} v^{k+1}_J$
  \State define $ V^{k+1}_{Ip} = v^p_I $, $p=1,...,k+1 $
  \parState{construct tridiagonal $ H^{k+1}_{pq} $ 
     with diagonal elements equal to $ (\alpha_1,..., \alpha_{k+1}) $ 
     and super- and sub-diagonal elements equal to $ (\beta_1,..., \beta_{k})$}
  \parState{compute $ \bm{x}^{k+1} 
               = \lVert \bm{\hat{W}}_\omega \rVert 
                     \bm{V}^{k+1} \sqrt{\bm{H}^{k+1}} \bm{e}^0 $, 
     \\ with $ e^0_{1}=1 $ and $ e^0_{q} = 0 $, $ q=2,...,k+1 $}
  \State set $ \Delta = \lVert \bm{x}^{k+1} - \bm{x}^{k} \rVert$
  \State set $ k=k+1 $
  \EndWhile
  \State set $ \hat{F}_{I,\omega} = x^{k}_I$
  \EndFor
  \State compute $ F_{I,n} = \frac{1}{\mmax}
    \left( \hat{F}_{I,0}+2\sum_{\omega=1}^{\mmax-1} \hat{F}_{I,\omega} \right)$ 
 \end{algorithmic}
\label{alg:coloured_noise}
\end{algorithm}

\begin{itemize}

	\item If a clear time scale separation between the typical diffusion
times of the coarse-grained particles and the time scale of the memory can be
assumed, then Eq.~(\ref{eq:stochastic_force2}) simplifies to
\begin{equation}
\left\langle \bm{F}^\text{R}_{i,n+m} \bm{F}^\text{R}_{j,n} \right\rangle 
  = k_\text{B} T a_m \bm{K}^\text{p}_{m}(\bm{R}_{ij,n}) / \Dt.
\end{equation}
This enables, among other things, the precalculation of the Fourier transform 
of the memory kernels (step 2).

	\item In step 19, only the correlated random numbers $F_{I,n}$ are
determined and not the entire inverse Fourier transform of
$\hat{F}_{I,\omega}$. If one can assume time scale separation as discussed 
in the previous comment, one could further reduce the computing time
by simultaneously calculating correlated random numbers $F_{I,n+m}$ for a 
whole time window $m \in [0, m_\text{FT}-1]$. This would speed up the 
algorithm by a factor close to $m_\text{FT}$. 

	\item If the dynamics of the system is described sufficiently
accurately by single-particle self-memory kernels only, one can use the
algorithms presented in Refs.~\citen{Barrat2011, Jung2017}. In that case, the
parameters for the noise calculation can be precalculated before the simulation
run, which significantly decreases the computational costs.

\end{itemize}


\section{Reconstruction of distance-dependent memory kernels}
\label{sec:reconstruction}

Having established an efficient integrator for \gj{GLD} simulations, we will now
discuss the problem how to construct coarse-grained \gj{GLD} models from simulations
of fine-grained models. In a previous publication \cite{Jung2017}, we have
introduced a class of iterative memory reconstruction (IMR) methods, which
allows to determine memory kernels iteratively from known dynamic
correlation functions. In that work, the method was tested on systems with a
single CG particle only. Here, we will show how the method can be applied to
large systems with many particles, where two-particle contributions to the
memory kernel tensor become important. 

Specifically, we consider the short-range frequency-dependent hydrodynamic
interactions between nanocolloids in dispersion. The underlying fine-grained
molecular dynamics (MD) model consists of a Lennard-Jones (LJ) fluid and
hard-core nanocolloids of radius $3 \sigma$ at temperature $\kB T=\epsilon$.
Here and in the following, all quantities are given in units of the LJ diameter
$\sigma$, the LJ energy $\epsilon$ and the time $\tau = \sigma
\sqrt{m/\epsilon}$ ($m$ is the mass of LJ particles). More details on the model
and the simulations are given in Appendix \ref{sec:model}.  Since the system
has already been studied extensively in Refs.~\citen{Jung2017, Jung2017a}, we
refer to these references for an in-depth analysis of the dynamical correlation
functions. 

In the present work, we use these correlation functions as input to construct
the coarse-grained model. The reconstruction is performed in two steps. First,
the self- and pair-memory kernels are initialized, using the same procedure as
introduced in Ref.~\citen{Jung2017a}. For this purpose, we transform the GLE
for two particles at given distance into two single-particle GLEs that can
be integrated very efficiently. The result of this reconstruction is a
coarse-grained model that is valid for very dilute systems and ignores
many-body effects. In a second step, these many-body effects are included in
the coarse-grained model using the IMR technique \cite{Jung2017} in combination 
with \gj{GLD} simulations.

\subsection{Constructing non-Markovian coarse-grained models 
in the highly dilute limit}
\label{ssec:dilute}


If the two-particle contributions to the memory kernels are not affected by
multibody effects, one can efficiently determine them from the auto- and
cross-correlation functions using a set of effective one-dimensional GLEs.
This approximation corresponds to the highly dilute limit. To derive the
coarse-graining scheme, we start with the \gj{GLD} equations, Eq.\ (\ref{eq:GLE})
with (\ref{eq:GLD_basic}) and (\ref{eq:GLD_basic2}), for two particles at roughly
constant distance $R$ in the absence of a conservative force, $ \FFC(t)= 0$.
The equation for particle 1 is given by
\begin{align}
M \dot{{V}}_1(t)= &- \int_{0}^t \ud s \: \big( \Ks(t-s) 
    + \DKs(R,t-s) \big) {V}_1(s)\nonumber\\
&-\int_{0}^t \ud s \:  \Kpair(R,t-s) {V}_2(s)  + \partial {F}_1(t),
\label{eq:gle1}
\end{align}
and the equation for particle 2 is analogous. Here, only two-particle memory
contributions that are parallel to the line-of-centers between the particles
are taken into account (see the discussion at the end of Sec.\
\ref{ssec:GLD_basic}). Hence, it suffices to consider the one-dimensional
motion of the particles along that axis. 
The \gj{GLD} equations for $V_{1,2}(t)$ can be decoupled by considering the sum 
and the difference of the velocities,
$V_{\pm}(t) = \sqrt{\frac{1}{2}} (V_1(t) \pm V_2(t))$, giving
\begin{equation}
\label{eq:gle2}
M \dot{V}_\pm  = - \int_0^t \ud s \: K_\pm(R, t-s) V_\pm
+ \partial F_\pm
\end{equation}
with $K_\pm(R,t) = \Ks(t) + \DKs(R,t) \pm  \Kpair(R,t)$. 

The above equation can be transformed into a noise-free differential equation
for the velocity auto- and cross-correlation functions by multiplying
Eq.~(\ref{eq:gle2}) with $ V_\pm(0) $ and subsequently taking the 
\fs{ensemble} average,
\begin{equation}
M \dot{C}_{\pm}(R,t)=- \int_{0}^t \ud s \:  K_\pm(R,t-s) {C}_{\pm}(R,s).
\end{equation}

Here, we have introduced the additive and subtractive velocity correlation
functions $ C_\pm(R,t) = C_{11}(R,t) \pm C_{12}(R,t)$ with $C_{ij}(R,t) =
\langle V_i(t) V_j(0) \rangle$ (using $C_{11}=C_{22}$ and $C_{12}=C_{21}$).  On
the basis of these equations, the following coarse-graining procedure to
determine a first approximation for the distance-dependent self- and
pair-memory kernels is suggested:

{\em First}, we perform a MD simulation of freely
diffusing particles and calculate the velocity auto- and cross-correlation
functions,
\begin{align}
  C^\text{a}(R,t) &= \frac{1}{\mathcal{N}}\sum_{i, j\neq i}^{} 
    \left\langle  V_i(t) V_i(0) + V_j(t) V_j(0)  \right\rangle_{R_{ij}(t)=R},
\\
  C^\text{c}(R,t) &=  \frac{1}{\mathcal{N}} \sum_{i, j\neq i}^{} 
    \left\langle  V_i(t) V_j(0) + V_j(t) V_i(0)  \right\rangle_{R_{ij}(t)=R},
\end{align}
with $ \mathcal{N} =2 N (N-1) $. These functions describe the velocity
auto- and cross-correlations of particle $i$ in the vicinity of a particle 
$j$ at distance $R$. In the above definitions we assume that the distance is
approximately constant on the time scale on which the velocity correlation
functions decay. \fs{This is indeed the case in colloidal and nanocolloidal
systems, where the Brownian diffusion times and the Brownian relaxation times
differ by at least two orders of magnitude \cite{Jung2017a}}.
For numerical reasons, the correlation functions are binned
with a spatial discretization $ \Delta R $, similar to the binning that is used
for the memory kernels (see also next paragraph).

{\em Second}, we determine the additive and subtractive velocity correlation
functions $ C_\pm(R,t) $ and apply an iterative memory reconstruction (IMR)
scheme for the derivation of the memory kernels $ K_\pm(R,t) $. The fundamental
idea of IMR schemes is to successively adapt the memory kernels in a
coarse-grained simulation model in such a way that they perfectly reproduce the
target dynamical correlations of the underlying microscopic system.  This
technique was proposed in Ref.~\citen{Jung2017} and is strongly related to the
iterative Boltzmann inversion known from static coarse-graining
\cite{Reith2003}.  Here, we use a scheme that targets the velocity correlation
functions, i.e., an IMRV scheme. \fs{Numerically IMRV schemes were found to be
more accurate than IMRF schemes that target force correlation functions
\cite{Jung2017}}. We will discuss \fs{the specific IMRV scheme applied here} in
more detail in Sec.\ \ref{ssec:many_body}.  The only disadvantage of IMR
methods compared to other reconstruction methods \cite{B.Schnurr1997,
Fricks2009, Shin2010, Carof2014, Lesnicki2016, Lei2016} like the inverse
Volterra technique \cite{Shin2010} is an increase in computational effort,
since coarse-grained simulations have to be performed in each iteration step.
However, as discussed before, the coarse-grained simulations are not
time-consuming, since we only need to integrate two one-dimensional GLE
equations (\ref{eq:gle2}). 

{\em Third}, the self- and pair-memory kernels are calculated according to
\begin{align}
   \Ks(t) + \DKs(R,t) &= \frac{K_+(R,t) + K_-(R,t)}{2},\\
   \Kpair(R,t) &= \frac{K_+(R,t) - K_-(R,t)}{2}.
\end{align}
Subsequently, the two-particle contribution to the self memory, 
$\DKs(R,t) $ can be determined by subtracting the known single-particle 
memory $ K^\text{s}(t) $ (see Ref.~\citen{Jung2017}).  

The reconstruction in Step 2 can also be conducted using explicit inverse Volterra
techniques \cite{Shin2010} (as was done, e.g., in Ref.~\citen{Jung2017a}). For
the system considered in this work, however, the results obtained with this approach were much less accurate than those obtained with the
IMRV method. This was due to accumulating errors when applying the inverse
Volterra technique, which became most pronounced at large correlation times.
In our applications here, we found the IMRV method to be more robust
than other non-iterative memory reconstruction techniques
\cite{B.Schnurr1997, Fricks2009, Shin2010, Carof2014, Lesnicki2016, Lei2016}.

\begin{figure}
\includegraphics[scale=1.0]{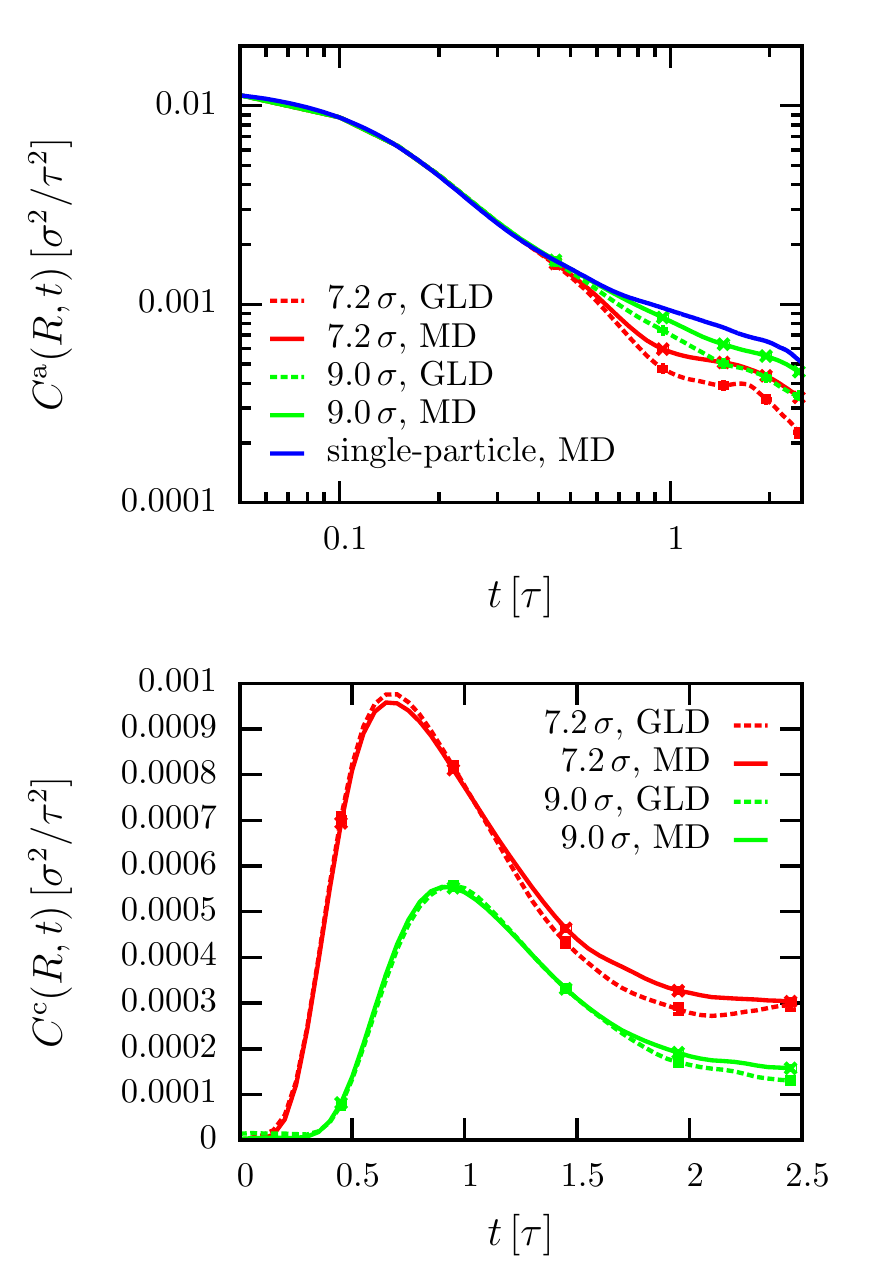}
\caption{Velocity auto- and cross-correlation functions ($\Ca$ and $\Cc$,
respectively) for pairs of nanocolloids at distances $R=7.2 \sigma$ (red lines)
and $R=9.0 \sigma$ (green lines) in a nanocolloid suspension with density $\rho
= 10^{-4} \sigma^{-3}$ (volume fraction 1 \%), as obtained from GLD
simulations (dashed) and MD simulations (solid). The input memory kernels in
the \gj{GLD} simulations are determined using the highly dilute approximation
described in Sec.\ \protect\ref{ssec:dilute}. Also shown for comparison is the
single-particle autocorrelation function of an isolated nanocolloid (with no
particles nearby) as obtained from MD simulations (blue solid line). To
visualize the statistical errors, five data points with y-error bars are
included in the curves. The error bars are smaller than the point size.}
\label{fig:GLDstep1}

\end{figure}

Fig.~\ref{fig:GLDstep1} compares the original results from the MD simulations
with those from \gj{GLD} simulations based on self- and pair-memory kernels
that were derived with the previously described coarse-graining procedure. The
velocity cross-correlation functions agree very well with each other. We can
thus conclude that many-body effects do not have a significant influence on the
cross-correlations in this system, at least for volume fractions around $ 1\%
$. Deviations between the \gj{GLD} and MD results are {observable}, however, in
the velocity auto-correlation functions. The velocity decorrelation is
significantly overestimated in the \gj{GLD} simulations for both distances $ R
$.  We explain this observation as follows: The comparison of the MD results
for single isolated nanocolloids with those for systems of many colloids
already shows that the velocity of a particle $i$ decorrelates more rapidly if
other particles are in its vicinity (see Fig.~\ref{fig:GLDstep1}, upper panel).
The reason is that the other particles disturb the flow field and thus reduce
the hydrodynamic backflow. When assuming the highly dilute limit, this
many-body effect is attributed to one {nearby} nanocolloid at distance $R$
only. Consequentially, the two-particle contribution to the self-memory,
$\DKs(R,t)$, is overestimated in the coarse-graining procedure, which leads to
the observed discrepancies. \gj{Additionally, the \fs{neglect} of the
conservative force in Eq.~\ref{eq:gle1} can also have an influence on the
observed discrepancies between the GLD and MD results as has recently been
pointed out in Refs.~\citen{Carof2014a} and \citen{Daldrop2017}. In our
simulations, however, we found that this effect was relatively small.}

To determine a coarse-grained model that accurately reproduces the MD results,
one must thus combine the iterative memory reconstruction with multiparticle
GLD simulations.

\subsection{Accounting for many-body effects in 
  non-Markovian coarse-grained models}
\label{ssec:many_body}

In this subsection, we will show how to apply the IMRV method for the
reconstruction of the frequency-dependent hydrodynamic interactions between
freely diffusing nanocolloids in many-body systems. In the IMR schemes, the
memory kernel $K(t)$ in the CG model is iteratively adjusted according to a
prescription \cite{Jung2017}
\begin{equation}
K_n \to K_{n+1} = 
    K_n + h_n(t) \big( \phi(Y_\text{MD}) - \phi(Y^{(n)}_\text{CG}) \big),
\end{equation}
such that differences of the target property, $Y$, in the CG model and in the
fine-grained MD model are successively reduced. Here $h_n(t)$ is a filter
function that changes from one iteration to the next (see Ref.\
\citen{Jung2017}), and $\phi(Y)$ is a mapping function, which must be optimized
depending on the choice of $Y$.  Specifically, the IMRV method targets the
velocity correlation functions.  In Ref.\ \citen{Jung2017}, a mapping function
based on the second time derivative of $Y(t)$ was proposed. However, when
applying this scheme to the present multiparticle system, we sometimes
observed an accumulation of errors in $Y_\text{CG}(t)$ at late times that
increased linearly with $t$. A better result was obtained with the alternative
mapping function
\begin{equation}
\phi(Y) = -\alpha \frac{M^2}{\kB T} \frac{Y(t+\Dt) -  Y(t)}{\Dt},
\label{eq:phi}
\end{equation}
where $ \alpha \approx 6\, \tau^{-1} $ is a relaxation parameter {that depends, i.a., on the nanocolloid density and the time step and needs to be adjusted for every reconstruction.}

We should mention that we focus in the present work on the hydrodynamic regime
in which at least a few solvent particles are located between the surfaces of
the colloids, i.e., we disregard lubrication forces in the \gj{GLD}
simulations.  This can be {implemented by applying} a lower cutoff $
r_\text{c,min} = 6.8\,\sigma $ to the memory kernels. For distances $ R<
6.8\,\sigma $, the memory kernels are replaced by the ones at $ R=
r_\text{c,min} $. The exclusion of lubrication forces is motivated by two
observations: First, for small distances, the statistics of the correlation
functions are worse than for larger distances, since the number of neighbors at
a distance $ R $ of a colloid scales quadratically with $ R $. Additionally,
the radial distribution function is significantly smaller than 1 in the regime
$ R < 6.8\,\sigma $.  Second, the lubrication forces lead to very strong pair
interactions between the particles, which contradicts some assumptions made in
this work. \gj{For example, in this case, the distance-dependence of the memory
kernel cannot be assumed to be negligible} on the time scale of the memory
$\tau_\text{mem} $. \gj{The \fs{neglect} of lubrication forces can of course have
an influence on the overall rheological properties of the system (as has, e.g.,
been discussed recently in Ref.~\citen{Ellero2016}). However, their influence
on the long-range dynamics should be small. We are therefore confident
that this assumption has no influence on the dynamical correlation functions
reported in the present work. }

The static pair-potential between the colloids is determined in a separate
coarse-graining procedure using the iterative Boltzmann inversion (IBI)
\cite{Reith2003}. Details can be found in \gj{Appendix~\ref{sec:model} and} Ref.~\citen{Jung2018}, Ch.~5. Since
we disregard lubrication forces, the pair potential did not have a significant
influence on the dynamical features studied in this work. 

\begin{figure}
\includegraphics[scale=1.0]{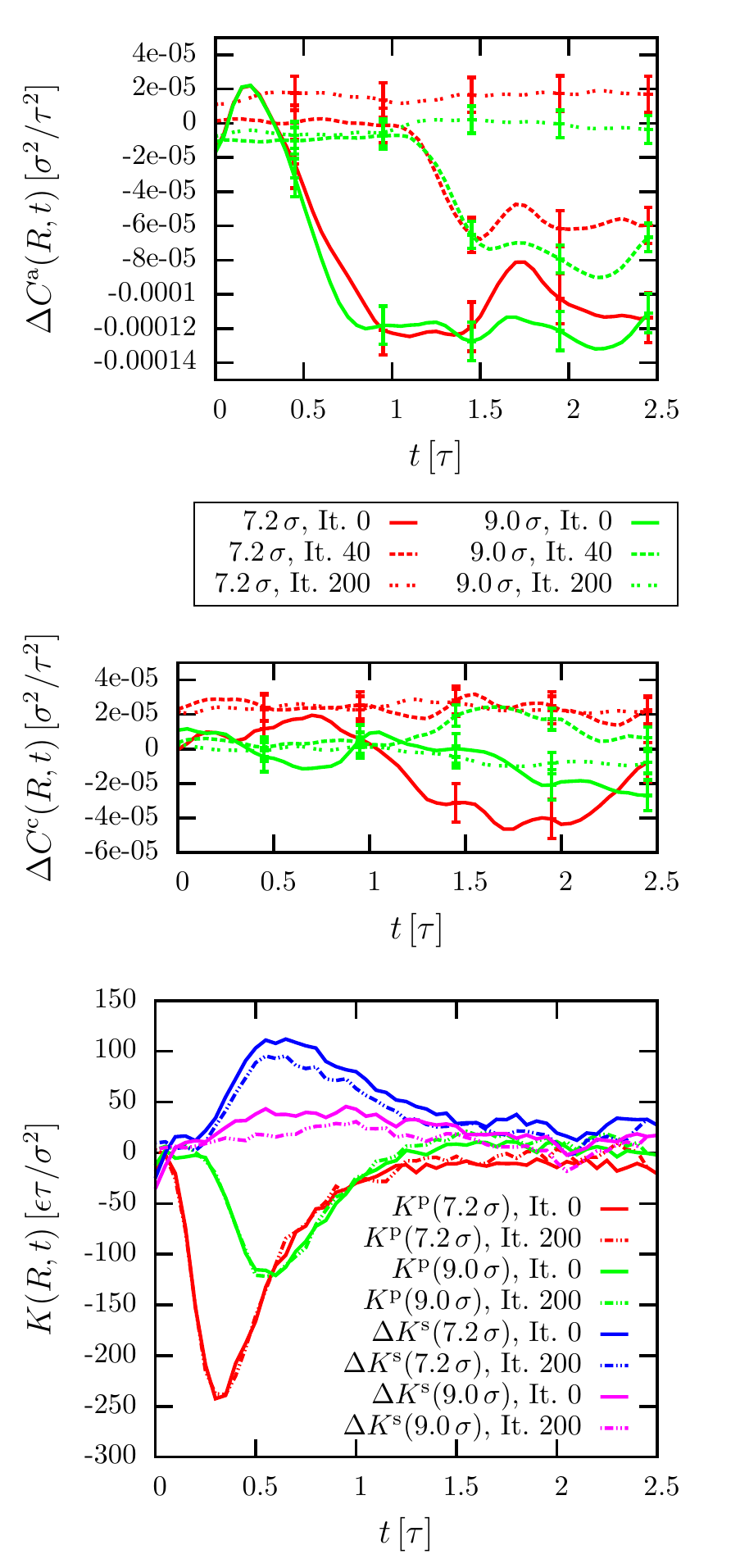}
\caption{Illustration of the iterative memory reconstruction (IMRV) method
applied to dispersions at nanocolloid density $\rho =
\rho_0:=10^{-4} \sigma^{-3}$. The memory kernels are initialized with the
results from the highly dilute approximation (Sec.\ \protect\ref{ssec:dilute}
and Fig.\ \protect\ref{fig:GLDstep1}).  The upper two panels illustrate the
evolution of the differences in the velocity auto- and cross-correlation
functions $\Ca$ and $\Cc$ between MD and \gj{GLD} simulations ($ \Delta C(t) =
C_\text{GLD}(t) - C_\text{MD}(t) $) for particle distances $R=7.2 \sigma$ (red)
and $R=9 \sigma$ (green) as the iteration progresses. The lower panel shows the
corresponding memory kernels. To visualize the statistical errors, five data
points with y-error bars are included in the curves.  } 
\label{fig:imrstep2}
\end{figure}

The results of the iterative procedure described in this subsection are shown
in Fig.~\ref{fig:imrstep2}. The reconstruction time step is $ \Delta t =
0.05\,\tau $ and \gj{the time scale of the correction in the iterative reconstruction} $ t_\text{cor} = 0.05 \,\tau $ \gj{(see Ref.~\citen{Jung2017} for details)}. Iterations $ 101-200 $
correspond to an iterative reconstruction initialized with the final memory
kernels of the first reconstruction (It.$~0-100 $). The two upper panels
visualize the iteration procedure. While the velocity cross-correlation
functions $\Cc$ are only subject to minor corrections if multiparticle effects
are taken into account, the autocorrelation function $\Ca$ is significantly
altered. In the last iteration step, the difference between MD and GLD
simulations is within the statistical errors, therefore, the iteration seems to
have converged.

The lower figure compares the initial memory kernels determined in the highly
dilute limit to the final results of the IMRV. It shows that the two-particle
contribution to the self-memory, $ \DKs(R,t) $, is significantly reduced in
dense systems. This is consistent with the previous discussions and
demonstrates that the multiparticle coarse-graining procedure is indeed capable
of incorporating many-body effects into the coarse-grained model, which were
neglected when assuming the highly dilute limit. The figure also shows that,
despite the reduction of $ \DKs(R,t) $, the distance-dependent correction to
the single-particle self-memory kernel still considerably differs from zero
and should not be ignored.

In sum, the results show that we can indeed reconstruct a dynamic
coarse-grained model for nanocolloids in dispersion that can reproduce the
dynamics of the underlying fine-grained system. 


\section{Analysis of the Non-Markovian Coarse-Grained Model}
\label{sec:results}

In this section we perform an in-depth analysis of the coarse-grained \gj{GLD} model
that was introduced in the previous section. We will show comparisons to MD
simulations and theory and analyze the transferability and efficiency of
the coarse-grained simulations.

\begin{figure}
	\includegraphics[scale=1.0]{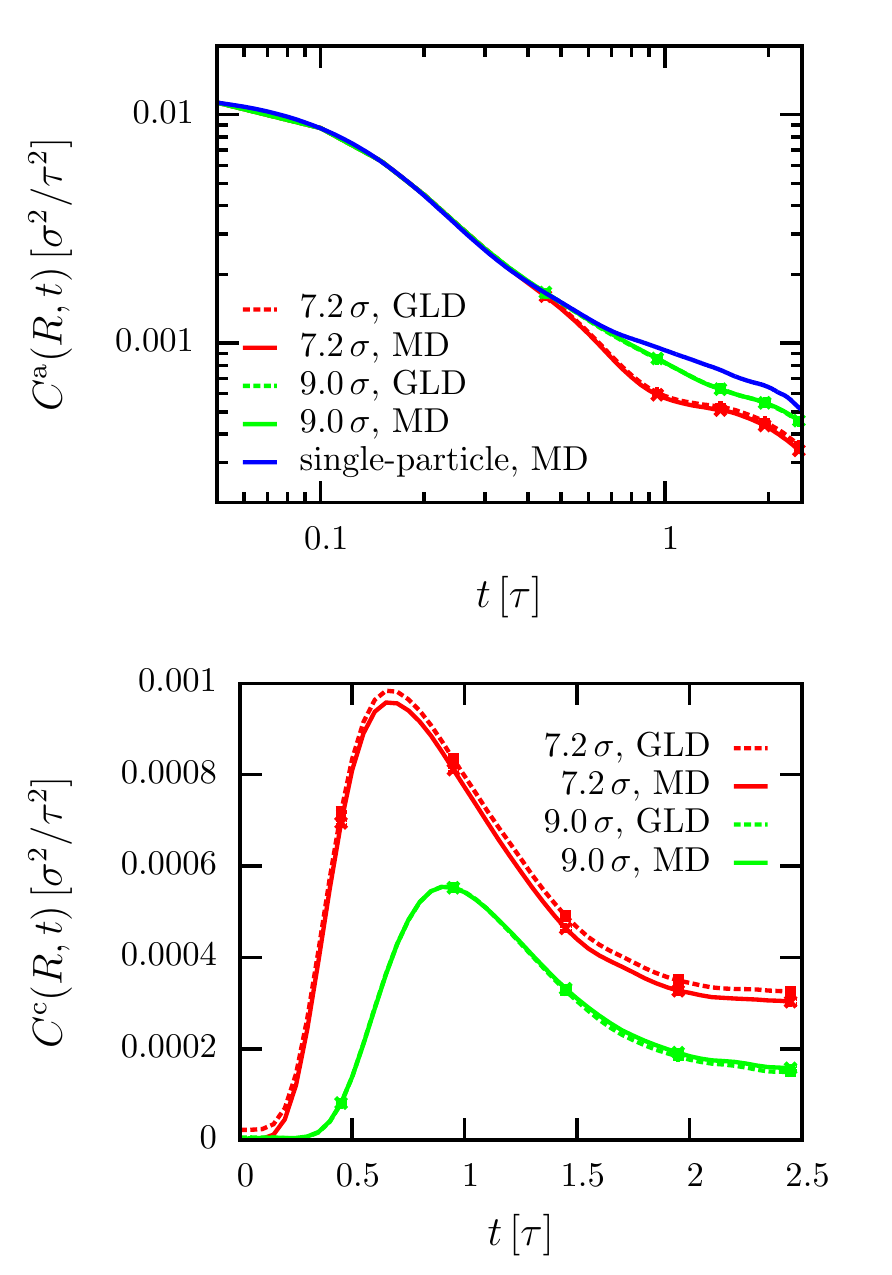}
\caption{Same as Fig.\ \protect\ref{fig:GLDstep1},
but now with \gj{GLD} memory kernels obtained by full many-body IMRV
reconstruction from MD simulation data at the same nanocolloid density 
$\rho=\rho_0 = 10^{-4} \sigma^{-3}$ (see Sec.\ \protect\ref{ssec:many_body} 
and Fig.\ \protect\ref{fig:imrstep2}).}
\label{fig:result1_c}
\end{figure}

\begin{figure}
	\includegraphics[scale=1.0]{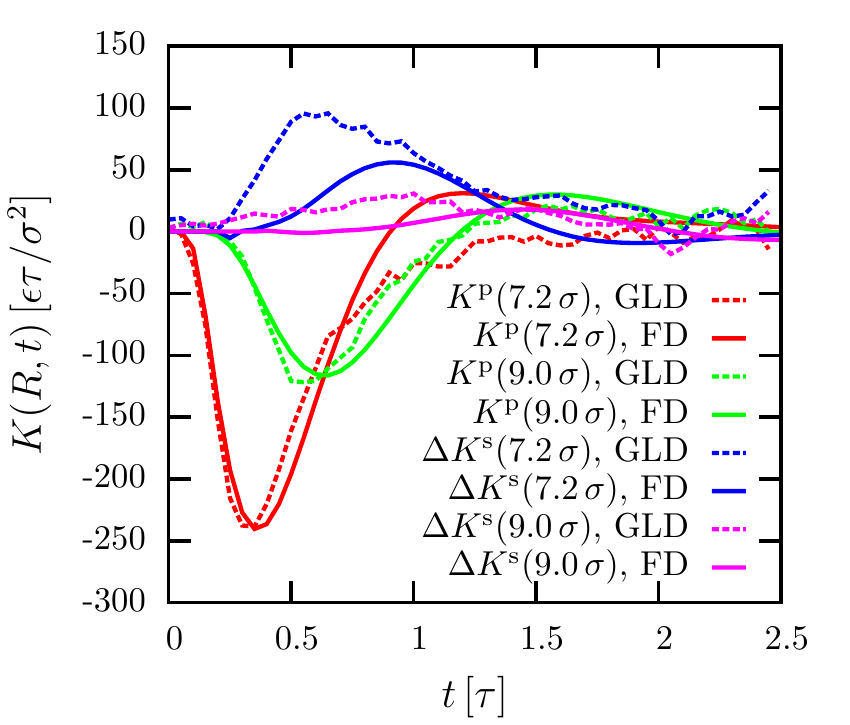}
\caption{Two-particle contributions to the memory kernels 
for different particle distances $R=7.2 \sigma$ and $R=9.0 \sigma$
as obtained from MD simulations via many-body IMRV reconstruction
(dashed lines, same data as in Fig.\ \ref{fig:imrstep2}) compared
to the analytical results in Ref.\ \citen{Jung2017a} from fluid 
dynamics (FD) theory. \fs{ As predicted by the theory, the time
scale of the peaks is $R/c_s$, where $c_s$ is the speed of sound.
}}
\label{fig:result1_k}
\end{figure}

\subsection{Comparison to MD simulations and hydrodynamic theory}

The dynamical properties of the system, namely the velocity auto- and
cross-correlation functions, are shown in Fig.~\ref{fig:result1_c}. The results
are compared to the dynamic correlation functions of the underlying microscopic
model. As expected from the well-behaved convergence of the reconstruction
procedure reported in the previous section, the differences between \gj{GLD} and MD
simulations are small. In fact, deviations can only be seen for small
particle distances, $ R < 8.0\,\sigma $.  The reason is that these
small-distance correlation functions have large statistical errors which
complicates the iterative memory reconstruction. Since the IMRV algorithm as
applied here (Eq.\ (\ref{eq:phi})) only corrects for differences in the first
derivative of the velocity correlation functions, it cannot easily handle
constant offsets. In order to improve on this, one would have to include a term
that depends on the target function $Y$ itself in the mapping function
$\phi(Y)$. {Alternative mapping functions, however, turned out to be less robust and since the deviations between \gj{GLD} and MD simulations are small we used the previously introduced mapping function Eq.~(\ref{eq:phi}).}

%
%
%
%

Fig.~\ref{fig:result1_k} compares the reproduced two-particle contributions to
the self- and pair-memory kernels to analytic results from hydrodynamic theory
\cite{Jung2017a}. In the case of the pair-memory kernel, the agreement is very
good. Deviations are only observed for small particle distances $R < 8.0
\sigma$, and they can be attributed to the approximate nature of the theory at
small distance to radius ratios (see discussion in Ref.\ \citen{Jung2017a}).
In contrast, significant differences between theory and simulations can be
noticed in the two-particle contribution to the self-memory kernel, $ \DKs(R,t)
$: The self-memory is substantially underestimated by the theory.  This could
have two reasons: First, $\DKs(R,t)$ strongly depends on the boundary condition
at the surface of the colloids, since it is caused by the reflection of sound
waves at nearby colloids. It is difficult to precisely model these reflections
with the hydrodynamic theory. Second, as discussed in the previous section, 
$\DKs(R,t)$ is affected by many-body effects which are not considered in the
theory.

\subsection{Transferability of memory kernels}
\label{secGLDa:transfer}

\begin{figure}
\centering\includegraphics[scale=1]{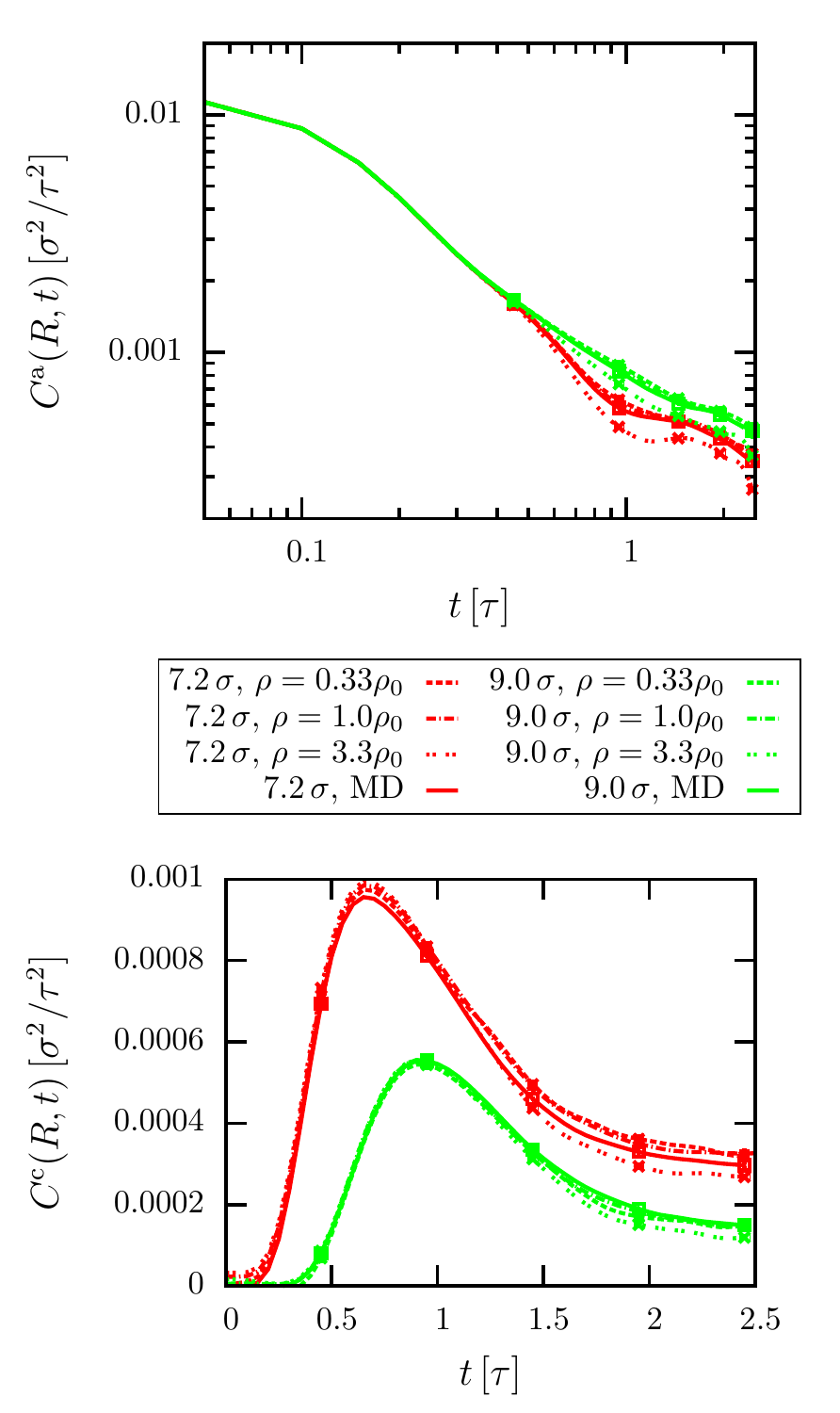}
\caption{Same as Fig.\ \protect\ref{fig:result1_c}, but now at different
nanocolloid densities $\rho$. In the \gj{GLD} simulations (dashed, dot-dashed and
dotted lines), the memory kernels derived in Sec.\ \protect\ref{ssec:many_body}
from MD simulations at the nanocolloid density $\rho_0 = 10^{-4} \sigma^{-3}$
are used (Fig.\ \protect\ref{fig:result1_k}). In the MD simulations, the
correlation functions do not visibly depend on the density, hence only one
representative curve (at $\rho = \rho_0$) is shown. }
\label{figGLDa:result4}
\end{figure}

In the following, we study the influence of the nanocolloid density $\rho$ on
the results for the velocity correlation functions.  This allows us to analyze
the transferability of the memory kernels determined in the previous section.
All \gj{GLD} simulations in this section are based on the memory kernels determined
by full many-body IMRV from reference MD simulations at density $\rho=\rho_0$.
The results are shown in Fig.~\ref{figGLDa:result4}. In the fine-grained MD
system (solid line), the effect of the density on the velocity correlation
functions is negligible and not visible in the plot. Similarly, a {\em
reduction} of the nanoparticle density in the \gj{GLD} simulations only has a minor
influence on the velocity autocorrelation functions.  A significant {\em
increase} of the nanocolloid density, however, leads to substantial differences
in the correlation functions in the \gj{GLD} model, which are {thus} not compatible with
the MD data. This can presumably be explained by the form of the self-memory
kernel in Eq.\ (\ref{eq:GLD_basic2}).  According to this equation, the
self-memory is a sum of the single-particle contribution, $\KKs$, and the
two-particle contributions, $\DKKs$, of all particles that are within the
cutoff $\rcut$ of the memory kernels.  If a particle $i$ is surrounded by many
other particles, however, the simple sum of two-particle contributions which
have been determined at a much smaller reference density is no longer
appropriate.  As discussed in Sec.\ \ref{ssec:many_body}, $\DKKs$ also includes
many-body effects in an effective manner and should thus depend on the density.

The discrepancies between \gj{GLD} and MD results at $\rho=3.3 \rho_0$ shown in
Fig.\ \ref{figGLDa:result4} are still comparatively small, but they increase
further as $\rho$ increases. In general, the memory kernels determined at the
density $\rho_0 = 10^{-4} \sigma^{-3}$ seem to be transferable to \gj{GLD}
simulations in the range of densities up to $\rho \lesssim 5 \rho_0$. At even
higher densities, $\rho \gtrsim 10 \rho_0$, an additional problem emerges: In
the simulations, one increasingly often encounters configurations {for which}
the total memory tensor $\Full{K}$ is not positive definite. This indicates a
stability problem, and moreover, the algorithm (Sec.\ \ref{ssec:GLD_RNG}) to
construct stochastic forces that satisfy the fluctuation-dissipation theorem
fails. In this region the assumption made in Eq.~(\ref{eq:GLD_basic2}) breaks
down and a more sophisticated many-body description has to be found. \gj{
Simply introducing density-dependent memory kernels would raise similar questions 
on the applicability and interpretation of these memory kernels as have been
discussed for many years for the case of static pair-potentials (see e.g.
Ref.~\citen{Louis2002}). }



%
%

With the above analysis, we have demonstrated that the non-Markovian
coarse-grained model derived in the previous section is applicable to a large
range of nanocolloid densities $ \rho $.  The reconstructed memory kernels thus
indeed characterize the fundamental hydrodynamic self- and pair-interactions
between the nanocolloids. Although this sounds obvious, it is an important
conclusion: By applying the iterative memory reconstruction, one can construct
coarse-grained models with dynamical features that perfectly match the
underlying fine-grained system, without necessarily capturing the fundamental
physical features of the system. This would be the case, e.g., if the set of
relevant dynamical variables chosen for the reconstruction of the
coarse-grained model were not ideal. Since the Mori-Zwanzig formalism does not
give any guidance regarding the optimal choice of relevant variables, this
decision has to be made \emph{a posteriori} and appropriately analyzed and
discussed \cite{Zwanzig2001}.

\subsection{Benchmarks of the \gj{GLD} simulations}

The potential benefits of \gj{GLD} simulations of course strongly depend on the
computational costs of the simulations. We have already discussed that the
algorithm scales linearly with the number of particles. Here, we will benchmark
the method more quantitatively (see Table~\ref{tabGLDa:benchmarks}). 

Not surprisingly, the simulation time of the fine-grained system for fixed
nanocolloid number ($N=125$) strongly decreases with increasing nanocolloid
density: The system contains fewer solvent particles per nanocolloid if the
density goes up. In contrast, the computational costs of \gj{GLD} simulations
increases with increasing density, since the nanocolloids have more neighbors
within their interaction range. Nevertheless, Table \ref{tabGLDa:benchmarks}
shows that the computation time per simulation step is still significantly
smaller in the \gj{GLD} simulations than in the fine-grained MD simulations in all
situations considered here. Furthermore, the time step in the \gj{GLD} simulations
is larger, which further speeds up the simulations. In total, the speedup of
the systems considered in this work ranges from $ S_\text{u} = 1017 \Dt[GLD] /
\Dt[MD] = 5.1 \cdot 10^4 $ (for $ \rho = 0.33\cdot 10^{-4}\,\sigma^{-3}$,
corresponding to a nanocolloid volume faction of  0.33 \%) to $ S_\text{u} = 44
\Dt[GLD] / \Dt[MD] = 0.2 \cdot 10^4  $ (for $ \rho = 3.3\cdot
10^{-4}\,\sigma^{-3}$, i.e., volume fraction 3.3\%). It is also interesting to
compare the costs of \gj{GLD} simulations with the simulation costs of a
corresponding Markovian implicit-solvent model without memory kernels (labeled
CG MD in Table~\ref{tabGLDa:benchmarks}). The \gj{GLD} simulations are slowed down
by a factor of roughly $150$ compared to the Markovian simulations.  This
slowdown is the prize one has to pay for including the correct dynamics
according to the generalized Langevin equation.  Nevertheless, the GLD
simulation technique still permits to speed up simulations of nanocolloid 
dispersions by at least 3 orders of magnitude while capturing the correct
dynamics.

\renewcommand{\arraystretch}{1.1}
\begin{table}[t]
\centering\begin{tabular}{cccc}
  \toprule 
    & \multicolumn{3}{c}{$\rho\,[\sigma^{-3}]$}\\ 
  \cmidrule[0.4pt](lr{0.125em}){2-4}
     Simulation method & 
        $0.33\cdot 10^{-4}$ & $1.0\cdot 10^{-4}$ & $3.3\cdot 10^{-4}$ \\ 
   \cmidrule[0.4pt](r{0.125em}){1-1}
   \cmidrule[0.4pt](lr{0.125em}){2-2}
   \cmidrule[0.4pt](lr{0.125em}){3-3}
   \cmidrule[0.4pt](l{0.25em}){4-4}
   \myrowcolour
    MD & $61000 \pm 2000\, $s & $ 19500 \pm 750\, $s & $  6400 \pm 100\, $s \\ 
		
   \gj{GLD} & $ 60 \pm 1\, $s & $ 63 \pm 1\, $s & $ 145 \pm 1\, $s\\ 
    \myrowcolour
   CG MD & $ 0.35 \pm 0.01\, $s & $ 0.56 \pm 0.01\, $s & $ 1.19 \pm 0.02\, $s\\ 
    \bottomrule
\end{tabular} 

\caption{Benchmarks of the \gj{Generalized Langevin dynamics} (GLD) technique
compared to molecular dynamics (MD) simulations for different nanocolloid
densities $ \rho $. The coarse-grained (CG) MD results refer to simulations
using only the mean conservative force $ F^\text{C}(R) $. The simulations
correspond to 125 nanocolloids integrated for $ 10^4 $ time steps, with time
step $ \Delta t_\text{MD} = 0.001\,\tau $ and $ \Delta t_\text{GLD,CG-MD} =
0.05\,\tau $. The benchmarks were run on the Mogon I cluster
(\url{https://mogonwiki.zdv.uni-mainz.de/dokuwiki/nodes}) using 1 core on an
``AMD Opteron 6272'' CPU.}

\label{tabGLDa:benchmarks}
\end{table}
\renewcommand{\arraystretch}{1.0}

\subsection{Final discussion and remarks}
\label{ssec:final}

The methods proposed in this work rely on the introduction of a cutoff in the
memory functions in both time and space. Since hydrodynamic interactions are
long-range and have long-time tails, it is thus not possible to reproduce them
in full detail with the reconstructed non-Markovian models. Methods that are
able to implement long-range interactions are often based on reciprocal space
calculations like the Ewald summation \cite{Ewald1921}. Due to the additional
time-dependence of the memory kernels, this technique cannot be adapted to the
generalized Langevin equation in a straightforward manner.  

One possible way to include long-range interactions in the model could be to
combine the generalized and the standard Brownian dynamics (BD) technique
\cite{Ermak1975,VanGunsteren1982}.  The former is then used to model the
frequency-dependent short-range interactions, $ R < \rcut $, between the
particles, and the latter accounts for long-range interactions at  $R > \rcut$,
using Ewald summation. In the most naive implementation of this scheme,
the pair memory term $\KKpair[\RR,\tau]$ would jump discontinuously 
from $\KKpair[\RR,\tau]= \KKpair[\text{GLD}](\RR,\tau)$ at $R < \rcut$
to $\KKpair[\RR,\tau] = \KKpair[\text{BD}](\RR) \: \delta(\tau)$
at $R > \rcut$. It may also be possible to link the two regimes in 
a continuous manner, if it is possible to find a function $\Phi(R)$ with
$\Phi(\rcut)=1$ such that {$ \KKpair[\RR,\tau] 
= \KKpair[\text{GLD}](\RR,\tau)|_{|\RR|=\rcut} \Phi(R)$,}
{with }
{
$ \Phi(R) \int_0^\infty \ud \tau \KKpair[\text{GLD}](\RR,\tau)|_{|\RR|=\rcut}
  = \KKpair[\text{BD}](\RR)$ for $R > \rcut$.} 
With such an Ansatz, the \gj{GLD} technique and Ewald summation methods
could possibly be combined.

The most important aspect of the non-Markovian \gj{GLD} technique introduced in the
present work is that it precisely models the sound waves that mediate the
hydrodynamic interaction between two nanocolloids.  When performing standard
Brownian dynamics simulations with instantaneous hydrodynamic interactions, two
approaching nanocolloids interact via an increasing friction force that
decelerates the particles. The time correlation function between the particles
decays rapidly. In contrast, the self- and pair-memory kernels in the
non-Markovian system generate time-delayed interactions due to the finite
propagation velocity of the sound waves. The memory-kernels thus induce a
long-time correlation between the velocities of the particles, which has a
positive sign, indicating that the particles move in the same direction (see
Fig.~\ref{fig:result1_c}, bottom panel). An important consequence of this
correlation is that two particles are less likely to approach each other. This
effect is captured by \gj{GLD} simulations, but not by Markovian BD simulations.
Additionally, the sound waves are reflected and interact with other colloids,
which also influences the collective dynamics and is not captured by BD
simulation techniques. The coarse-graining techniques developed in the present
paper enable us to systematically investigate such memory effects and
incorporate them into non-Markovian coarse-grained models.  The methods thus
open up new ways for dynamical coarse-graining and for understanding dynamical
processes in soft matter physics.


\section{Conclusion}
\label{sec:conclusion}

In this work we have introduced the ``\gj{Generalized Langevin dynamics}'' (GLD)
technique, a generalization of the Brownian dynamics (BD) method that includes
time-dependent friction kernels. We have applied the \gj{GLD} technique to study the
frequency-dependent hydrodynamic interactions between two nanocolloids in a
compressible fluid. The diffusion of these nanocolloids is affected by the
hydrodynamic backflow effect, i.e., the movement of the colloids induces fluid
vortices that interact with themselves and other colloids at later times.
Additionally, these colloids have pronounced pair-interactions mediated by
sound waves. Combining the iterative memory reconstruction technique introduced
in Ref.~\citen{Jung2017} with \gj{GLD} simulations, we can reconstruct non-Markovian
coarse-grained models that precisely reproduce these transversal and
longitudinal dissipative interactions. The model has a speedup of up to $
S_\text{u}=10^4 $ compared to MD simulations of the underlying fine-grained
system, and it is transferable to a wide range of nanocolloid densities.
Additionally, we have shown that the reconstructed memory kernels are
quantitatively comparable to those predicted from fluid dynamic
theory~\cite{Jung2017a}. 

The method can be applied without further adjustments to systems where
long-range dissipative interactions are not important, such as
frequency-dependent dipole interactions or active particles with
non-instantaneous local reorientations. In cases where long-range hydrodynamic
interactions cannot be neglected, it could be combined with standard BD
techniques as discussed in Sec.\ \ref{ssec:final}. 

An interesting topic for future research would be the construction of an
auxiliary variable expansion that describes the same generalized Langevin
equation as considered in this manuscript. For example, it might be possible to
adapt methods from numerical analysis concerning the construction of hidden
Markov models to this problem. Such an approach might be more promising than
the \emph{a posteriori} fitting procedures to auxiliary variable expansions
that have been proposed in the literature, since the application of the latter
to distance-dependent memory kernels turned out to be difficult (see the
discussion in the Appendix~\ref{sec:aux_var}). 

To generalize our ideas to non-equilibrium systems, it will also be important
to obtain a deeper understanding of the appropriate form of the generalized
Langevin equations (GLEs) in non-equilibrium situations, e.g., using the
Mori-Zwanzig projection formalism\ \cite{Meyer2017}. In particular, the
fluctuation-dissipation theorem which defines the correlations of the
stochastic forces at equilibrium is \fs{not necessarily} valid at
non-equilibrium\fs{\cite{Meyer2017}}. \fs{For example, it will be violated, in
systems subject to active noise}.  Since the main focus of soft matter research
currently shifts from equilibrium to non-equilibrium systems, we expect that
increasing effort will be devoted to dynamic coarse-graining in the future, and
that many exciting new techniques and applications will emerge. We hope that
the methods presented here will prove useful in this future research.


\section*{Acknowledgment}

This work was funded by the German Science Foundation within project A3 of the SFB TRR 146. Computations were carried out on the Mogon Computing Cluster at ZDV Mainz.


\appendix

\section{Derivation of the \gj{GLD} integrator equations}
\label{sec:GLD_integration}

Our starting point is Eq.~(\ref{eq:GLE_discretized}) in the main text.
We proceed in several steps.
In the {\em first} step, we integrate Eq.~(\ref{eq:GLE_discretized})
from $ t_n = n\Delta t $ to $ t_{n+1}=t_n+\Delta t $. Without
applying any approximation, the result is
\begin{eqnarray}
\label{eq:DP}
M (\VV_{i,n+1} - \VV_{i,n})
&=& - \KKs[0](\RR_{i,n+1}-\RR_{i,n})
\\ \nonumber  
&& + \DPC[i,n] + \DPD[i,n] + \DPR[i,n],
\end{eqnarray}
where the first term singles out the momentum change of particle $i$ in the
time step $\Delta t$ due to the instantaneous single-particle friction
$\KKs[0]$, and the other terms summarize other contributions to the total
momentum change of particle $i$.  The single-particle friction term is singled
out, because it often dominates.  Therefore, it will incorporated in a
semi-implicit integration scheme in the second step of this derivation scheme.
The remaining dissipative contributions to the total momentum change are given
by
\begin{equation}
\DPD[i,n] = \DPDpp[i,n,0] 
    - \sum_{m=1}^{\mmax - 1} \Big( \KKs[m] \Delta \RR_{i,n-m} 
               + \DPDpp[i,n,m] \Big)
\end{equation}
with $\Delta \RR_{k,n} = (\RR_{k,n+1} -\RR_{k,n})$ and
\begin{eqnarray}
\DPDpp[i,n,m] &=& \sum_{j \neq i} 
  \hspace*{-0.2cm} \int\limits_{t_n}^{t_n + \Dt} \hspace*{-0.2cm}
   \ud t\: \Big(\KKpair[m][\RR_{ij}(t)] \, \VV_j(t-m \Dt) 
\nonumber
\\ && \hspace*{0.5cm}
  + \DKKs[m] [ \RR_{ij} (t) ] \, \VV_i(t-m \Dt) \Big). 
\label{eq:dpdpp} 
\end{eqnarray}
The conservative contribution is
\begin{equation}
\DPC[i,n] = \int_{t_n}^{t_n + \Dt} \ud t \: \FFC[i][\{\RR_{j}(t)\}], 
\end{equation}
and the stochastic contributions $\DPR[i,n]$ are vectors of
random numbers with mean zero and correlations
\begin{eqnarray}
\nonumber
\langle \DPR[i,n+m] \DPR[j,n] \rangle &=&
\kB T \, a_m 
  \hspace*{-0.5cm}  \int\limits_{t_{n+m}}^{t_{n+m} + \Dt} \hspace*{-0.5cm} 
  \ud t \: \KKpair[m][\RR_{ij}(t)] \quad : \, j \neq i 
\\ 
\label{eq:stochastic}
\langle \DPR[i,n+m] \DPR[i,n] \rangle &=&
    \kB T \, a_m \KKs[m] \Dt 
\\ \nonumber && + \sum_{k \neq i} 
  \hspace*{-0.2cm} \int\limits_{t_{n+m}}^{t_{n+m} + \Dt} \hspace*{-0.5cm}  
  \ud t \: \DKKs[m][\RR_{ij}(t)]
\end{eqnarray}

In the {\em second} step, we construct a semi-explicit integration
scheme using the approximation
\begin{equation}
\RR_{i,n+1} - \RR_{i,n} = \frac{\Dt}{2} (\VV_{i,n+1} + \VV_{i,n})
 + {\cal O}(\Dt^3)
\label{eq:DR_appr}
\end{equation}
Combining (\ref{eq:DR_appr}) with (\ref{eq:DP}), we obtain the
following equations, which are accurate up to order ${\cal O}(\Dt^2)$.
\begin{eqnarray}
\RR_{i,n+1} &=& \RR_{i,n} + \Dt \bbb \: \VV_{i,n} 
\nonumber \\  && 
   + \frac{\Dt}{2M} \bbb \: (\DPC[i,n] + \DPD[i,n] + \DPR[i,n]) \:
\label{eq:rrstep}
\\
\label{eq:vvstep}
\VV_{i,n+1} &=& \aaa \VV_{i,n} 
  + \frac{1}{M} \bbb (\DPC[i,n] + \DPD[i,n] + \DPR[i,n]) \quad
\end{eqnarray}
where the matrices $\aaa$ and $\bbb$ are given by Eq.\ (\ref{eq:aa-bb}) in
the main text.
  
In the {\em third} step, we introduce further approximations for $\DPC[i,n]$
and $\DPD[i,n]$ in Eqs.\ (\ref{eq:rrstep}) and (\ref{eq:vvstep}).  The
conservative contribution, $\DPC[i,n]$, is approximated by
expressions such that the error in Eqs.\ (\ref{eq:rrstep}) and
(\ref{eq:vvstep}) remains of order ${\cal O}(\Delta t)^3$. In Eq.\
(\ref{eq:rrstep}), it is sufficient to use the approximation $\DPC[i,n] = \Dt
\FFC[i,n] + {\cal O}(\Dt^2)$ with $\FFC[i,n] = \FFC[i][\{\RR_j(t_n)\}]$ (see
also main text). In Eq.\ (\ref{eq:vvstep}), an expression of order ${\cal
O}(\Dt^2)$ is required, e.g., $\DPC[i,n] = \frac{\Dt}{2}(\FFC[i,n] +
\FFC[i,n+1]) + {\cal O}(\Dt^3)$.  Here, we introduce an additional correction
term of order ${\cal O}(\Dt^3)$ which does not change the order of the
algorithm, but restores the form of the quasi-symplectic ''SVVm'' algorithm of
Melchionna in the case of Markovian single-particle Langevin dynamics. Thus we
use (see also Ref.\ \citen{Gronbech-Jensen2012})
\begin{eqnarray}
\nonumber
\DPC[i,n] &=& \frac{1}{2} \Big(
(1-\frac{\Dt}{2M} \KKs[0]) \FFC[i,n]
+ (1+\frac{\Dt}{2M} \KKs[0]) \FFC[i,n+1] \Big)
\\
\Rightarrow \bbb \DPC[i,n] &=& \frac{\Dt}{2} (\aaa \FFC[i,n] + \FFC[i,n+1]).
\end{eqnarray}

Regarding the dissipative and stochastic contributions, $\DPD[i,n]$ and
$\DPR[i,n]$, we assume that the two-particle contributions $\KKpair[m]$ and
$\DKKs[m]$ vary sufficiently slowly as a function of $\RR_{ij}(t)$ so that they are approximately constant during one time step. Hence we make the approximation
\begin{equation}
\left.  \begin{array} {lcl}
\DKKs[m][\RR_{ij}(t)] &\approx& \DKKs[m][\RR_{ij,n}] \\
\KKpair[m][\RR_{ij}(t)] &\approx& \KKpair[m][\RR_{ij,n}]
\end{array} \right\} \; \mbox{for} \; 
t \in [t_n, t_n + \Dt]. 
\end{equation}
The integrals in Eq.\ (\ref{eq:stochastic}) are thus replaced by
$\int_{t_n}^{t_n + \Dt} \ud t \: \KK[m][\RR_{ij}(t)] \approx \Dt
\KK[m][\RR_{ij,n}]$, and the expression for $\DPDpp[i,n,m]$ (Eq.\
(\ref{eq:dpdpp})) is replaced by
\begin{eqnarray}
\DPDpp[i,n,m] &=& - \sum_{j \neq i} \Big(
  \KKpair[m][\RR_{ij,n}] \, \Delta \RR_{j,n-m} 
\nonumber \\  && \hspace*{1cm}
  + \DKKs[m][\RR_{ij,n}] \, \Delta \RR_{i,n-m} \Big) \qquad
\end{eqnarray}
for $m > 0$. However, in order to obtain an integrator based on explicit
equations for $\RR_{i,n+1}$ and $\VV_{i,n+1}$ in each integration step,
the instantaneous friction contribution $\DPDpp[i,n,0]$ must be
treated differently. Here, we make the approximation
\begin{equation}
\DPDpp[i,n,0] = - \Dt \sum_{j \neq i} \Big(
  \KKpair[m][\RR_{ij,n}] \VV_{j,n} + \DKKs[m][\RR_{ij,n}] \VV_{i,n}\Big)
\end{equation}
This completes the set of approximations that enter the algorithm.
Putting everything together, we obtain Eqs.\ (\ref{eq:GLD_integrator})-
(\ref{eq:stochastic_force2}) in the main text.

We note that in the present algorithm, the instantaneous single-particle
friction contributions ($\KKs[0]$) and the two-particle contributions
($\DKKs[0]$ and $\KKpair[0]$) are treated on different footings. This is
because we have assumed that $\KKs[0]$ dominates over the two-particle
contributions, hence we include only $\KKs[0]$ in the implicit part of our
semi-implicit scheme. Including $\DKKs[0]$ and $\KKpair[0]$ as well is
possible, but results in a scheme where high dimensional matrices have to be
inverted in each time step (this can be done efficiently, e.g., via the Lanczos
algorithm \cite{Hochbruck1997, Higham2008, Aune2013}). In the applications
considered in the present work (Sec.\ \ref{sec:results}), the instantaneous
two-particle contributions to the memory vanished, due to the fact that the
memory effects are mediated by propagating waves in the solvent. Hence the
approximation $\KKs[0] \gg \DKKs[0],\KKpair[0]$ was justified.

\section{Simulation details}
\label{sec:model}

In the fine-grained reference simulations, we consider systems of $N=125$
nanocolloids in a Lennard-Jones (LJ) fluid. The LJ particles have the mass $m$
and interact {\em via} a truncated and shifted LJ potential with LJ radius
$\sigma$, LJ energy amplitude $\epsilon$, and cutoff at $r_\text{c,LJ} = 2.5
\sigma$.  This corresponds to a hard-core interaction with a small attractive
tail.  The fluid is initialized by placing LJ particles on an fcc-lattice with
lattice constant $a=1.71 \sigma$ and therefore a density of $\rho = 0.8
\sigma^{-2}$.  Nanocolloids are then introduced into the system by overlaying
spheres of radius $R_c=3 \sigma$ onto the fcc-lattice, and then turning all LJ
particles inside the spheres into constituents of nanocolloids. This
results in nanocolloids of mass $M=80 m$.  Nanocolloids are rigid bodies, i.e.,
the relative distances of all particles forming one nanocolloid are kept fixed.
Particles that are part of a nanocolloid interact with other LJ particles by
purely repulsive interactions, i.e., {\em via} a truncated LJ potential with
cutoff $r_\text{c,LJ} = \sqrt[6]{2} \sigma$.  We use a cubic simulation box with
periodic boundary conditions in all three dimensions and box size $L^3=N/\rho$,
where $\rho$ is the target density of the nanocolloids.  The system is
equilibrated at the temperature $\kB T=\epsilon$ by $NVT$ simulations with a
Langevin thermostat.  Simulation data are then collected from 
molecular dynamics (MD) simulations in the $NVE$ ensemble with a time step
$\Dt[\text{MD}]=0.001 \tau$.  All fine-grained simulations are performed with
the simulation package {\em Lammps} \cite{Plimpton1995,Plimpton1995a}.  

Specifically, the memory kernels were reconstructed from nanocolloid
correlation functions in reference simulations of 125 nanocolloids with box
dimensions $L=107.728 \sigma$, corresponding to a number density $\rho_0 =
0.0001 \sigma^{-3}$ and a volume fraction of 1 \%.  The time step of the GLD
simulations is $\Dt=0.05 \tau$, and the cutoff of the memory sequence is
$\mmax=50$, corresponding to the memory scale $\tau_{\text{mem}}=2.5 \tau$.
The spatial discretization for the memory kernels is $\Delta R=0.2 \sigma$ with
a cutoff at $\rcut=15 \sigma$. From the mean thermal velocity of the
nanocolloids, $v = \sqrt{M^{-1}} \approx 0.1 \sigma/\tau$, one can conclude
that the particles roughly diffuse over a distance $\Delta R$ on the time scale
$\tau_{\text{mem}}$. Thus we may assume that two-particle contributions to the
memory kernels change only slightly on this time scale, which allows us to
optimize the \gj{GLD} algorithm by precalculating the Fourier transforms of the
memory kernels in Algorithm \ref{alg:coloured_noise} as described in Sec.\
\ref{ssec:GLD_RNG}. 

\begin{figure}
	\includegraphics[scale=1.0]{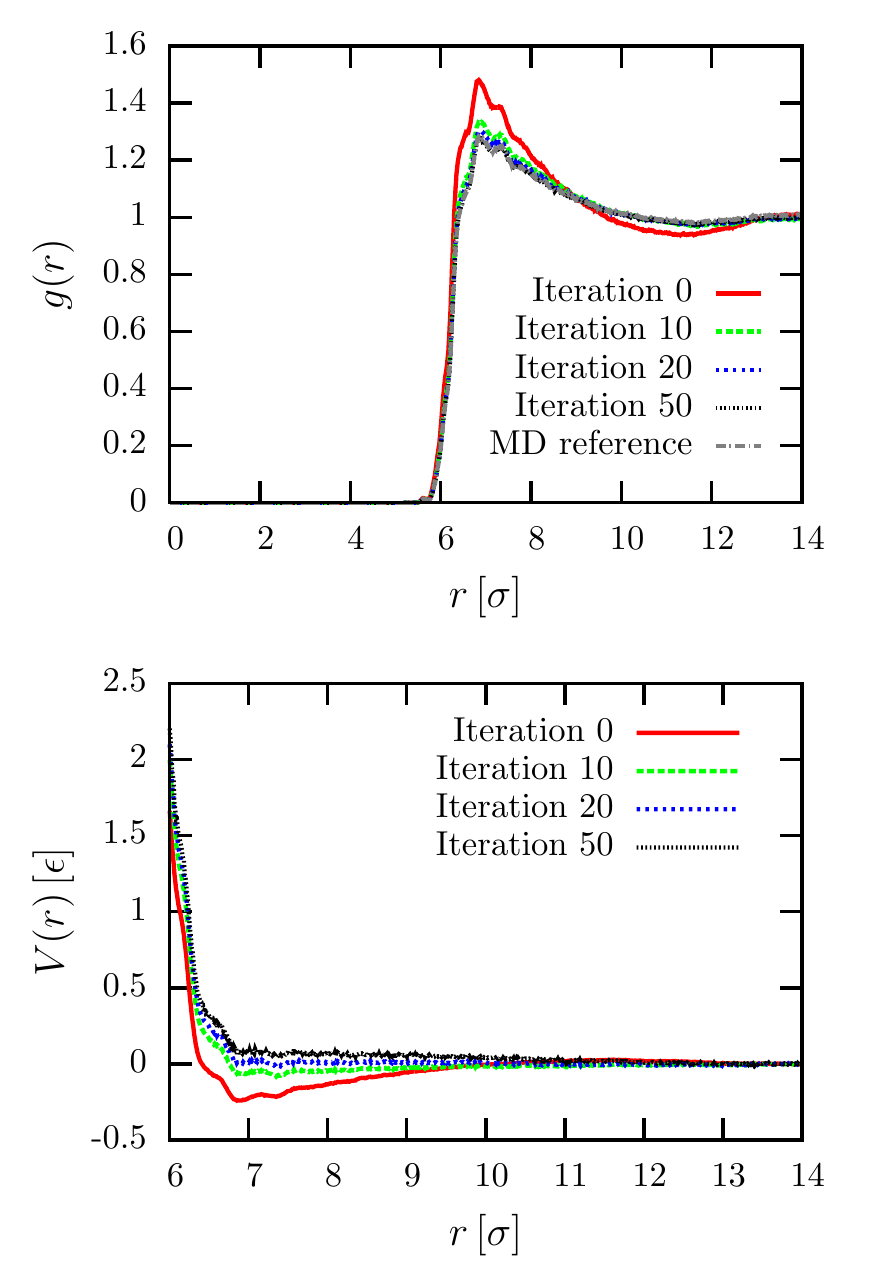}
	\caption{\gj{Iterative Boltzmann inversion (IBI) to determine the static pair-potential between two nanocolloids. The upper figure shows the different iterations of the radial distribution function $ g(r) $. The gray reference curve lies precisely on top of iteration 50. The lower figure visualizes the static pair potential $ V(r) $. For $ r<6\,\sigma $ the potential just corresponds to a hard-core potential.}}
	\label{fig:static_potential}
\end{figure}

\fs{Fig.~\ref{fig:static_potential} shows different steps of the 
iterative Boltzmann inversion (IBI)
reconstruction of the static pair-potential between two nanocolloids.
In the first step of the iteration (lower panel, iteration 0), one
simply makes the Ansatz that the potential is the negative logarithm
of the pair correlation function. At this level, it becomes negative
at intermediate distances. After applying the IBI procedure, however,
the resulting pair potential is purely repulsive. Hence the peak in
the radial distribution function (see upper panel) is due to layering 
of colloids and not due to solvent-mediated attractive interactions.}

\section{About the auxiliary variable expansion}
\label{sec:aux_var}

Most techniques that have been proposed so far to include non-Markovian
dynamics into coarse-grained models are based on an auxiliary variable
expansion, where the coarse-grained equations of motion are coupled to
additional variables with Markovian dynamics \cite{Ceriotti2010, Cordoba2012,
Cordoba2012a, Baczewski2013, Li2017}.  With such an approach, the non-Markovian
dynamics of the system can be reproduced, although the actual equations of
motion are purely Markovian.  The auxiliary variables are determined by fitting
(complex) exponential functions to the desired memory kernels. Unfortunately,
it was not possible to adapt this procedure to the simulations considered in
this work, mainly due to problems with the distance-dependent two-particle
contributions to the self- and pair-memory kernels. For future reference, we
shortly sketch the problems that occurred when attempting an auxiliary variable
expansion.

To replace the $N$-particle generalized Langevin equation~(\ref{eq:GLE}) 
with Markovian equations of motion, we use the Ansatz
\begin{align}
\begin{pmatrix} \dot{\bm{V}}(t) \\ \dot{\bm{s}}(t) \end{pmatrix} = \begin{pmatrix} \bm{F}^\text{C}(t) \nonumber \\ 0 \end{pmatrix} &- \begin{pmatrix} \bm{0} & \bm{A}^{vs} \\ \bm{A}^{sv} & \bm{A}^{ss} \end{pmatrix} \begin{pmatrix} \bm{V}(t) \\ \bm{s}(t) \end{pmatrix} \\
&+ \begin{pmatrix} \bm{0} & \bm{0} \\ \bm{0} & \bm{B} \end{pmatrix} \begin{pmatrix} \bm{0} \\ \zeta(t) \end{pmatrix} , \label{eq:approach}
\end{align}
with the auxiliary variables $ \bm{s}(t) $, the coupling matrices $ \bm{A}^{vs}
$, $ \bm{A}^{sv} $ and the dissipative matrices $ \bm{A}^{ss} $. We also
introduce uncorrelated Gaussian distribution random numbers $ \zeta(t) $ with
zero mean and unit variance. The matrices $ \bm{B} $ are given by the
fluctuation-dissipation theorem $ \bm{B} {\bm{B}}^\text{T} = \bm{A}^{ss} +
{\bm{A}^{ss}}^\text{T}$. The $ 2KN $-dimensional vector $ \bm{s}(t) $ consists
of $ K $ uncoupled auxiliary variables $ \bm{s}_k(t) $. In the following, the
subscript $ k = 0,...,K-1 $ will be used to refer to submatrices that act on
the auxiliary variable $ \bm{s}_k(t) $. Assuming that the time-dependence of
$\bm{V}(t)$ is known, the set of linear equations (\ref{eq:approach}) can be
solved. From this we can extract the time-evolution of the auxiliary variable
system,
\begin{equation}
\bm{s}_k(t) = \sum_{k=0}^{K-1} \int_{0}^{t} \text{d}t' e^{-(t-t')\bm{A}^{ss}_k} \left ( \bm{A}^{sv}_k \bm{V}(t') + \bm{B}_k\zeta_k(t') \right ).
\end{equation}
Inserting this result into the original approach shows that 
Eq.~(\ref{eq:approach}) indeed corresponds to a generalized Langevin equation,
\begin{equation}
M \dot{\bm{V}}(t)=  \bm{F}^\text{C}(t)-\sum_{k=0}^{K-1} \left( \int_{0}^{t} \text{d}s \bm{K}_k(t-s) \bm{V}(t) + \partial \bm{F}_k(t) \right).
\end{equation}
The memory kernel is given by
\begin{equation}
\bm{K}_k(t) = -M\bm{A}^{vs}_k e^{-t\bm{A}^{ss}_k} \bm{A}^{sv}_k,
\end{equation}
and the correlation function of the random force is,
\begin{equation}
\left \langle \partial \bm{F}_k(t) \partial \bm{F}_k(t') \right \rangle = k_\text{B} T M^2 \bm{A}^{vs}_k e^{-(t-t')\bm{A}^{ss}_k} {{\bm{A}^{vs}_k}^T}.
\end{equation}
To comply with the fluctuation-dissipation theorem, we thus set 
$ \mathbf{A}^{vs}_k = -M^{-1}{\mathbf{A}^{sv}_k}^T$. To simplify the notation, 
we also define
\begin{equation}
\mathbf{A}^{ss}_k = \begin{pmatrix} \mathbf{A}'^{ss}_k & 0 & 0 \\ 0 & ... & 0 \\ 0 & 0 & \mathbf{A}'^{ss}_k \end{pmatrix} \quad \text{with} \quad \mathbf{A}'^{ss}_k = \begin{pmatrix} q_k & r_k \\ -r_k & q_k \end{pmatrix},
\end{equation}
and introduce the reduced $ N \times N $-dimensional coupling matrices 
$ \bm{A}'^{sv}_{k} $,
\begin{align}
A^{sv}_{k,ij} &= A'^{sv}_{k,(2i)j}  \text{ for even }i,\\
A^{sv}_{k,ij} &= 0 \hspace*{0.985cm} \text{ for odd }i.
\end{align}
The latter definition transfers the complex exponential functions to scalar
exponential functions multiplied by a cosine. In general, it is also possible
to introduce an additional phase parameter to the memory kernel. In our system,
however, this complicates the fitting procedure without significantly
increasing the quality of the fits. These considerations finally lead to memory
kernels,
\begin{equation}
\mathbf{K}_k(t) = {\mathbf{A}'^{sv}_k}^T {\mathbf{A}'^{sv}_k} e^{-q_kt}\cos(r_kt) = \mathbf{P}_k e^{-q_kt}\cos(r_kt).
\end{equation}
The coupling matrices $ \bm{A}'^{sv}_k $ can be calculated from $ \bm{P}_k $ by
Cholesky decomposition. These fitting matrices $ \bm{P}_k $ contain the
information about the amplitudes of the exponential functions that are fitted
to the self- and pair-memory kernels. 

We tested this approach for the memory kernels that were reconstructed from the
fine-grained simulations. In these tests, we observed that including the self-
and pair-memory kernels resulted in fitting matrices $ \bm{P}^k $ that were not
positive definite.  Consequentially, the coupling matrices were not well
defined. This can be rationalized in the following way: While the memory
kernels themselves vary smoothly with the distances of particles, the fitting
procedure produced strongly discontinuous distance-dependent amplitudes.
These discontinuities could lead to impossible correlation matrices for some
auxiliary variables $k$, in which one particle was strongly correlated with the
other particles, while these other particles were anti-correlated. The problem
could be reduced by constraining the fitting parameters, which made the fitting
procedure more difficult without significantly improving the general
applicability of the technique. For our systems, we were thus not able to
construct a realistic and stable auxiliary variable expansion.

One possible solution for this problem could be to construct hidden
Markov models directly from the atomistic simulations and not based on an
\emph{a posteriori} fitting procedure \cite{Baum1966}. This, however, goes
beyond the scope of the present work.

\bibliography{library,library_local,library-control}

\end{document}